\input harvmac
\input epsf
\input amssym
\baselineskip 14pt



\def\d{\delta}
\def\db{\overline{\d}}
\def\pb{\overline{\partial}}

\def\cA{{\cal A}}
\def\cAb{{\cal \overline{A}}}
\def\psib{\overline{\psi}}
\def\Re{{\rm Re}}
\def\Im{{\rm Im}}
\def\wh{\widehat}
\def\Och{\widehat{{\cal O}}}
\def\Lb{\overline{L}}
\def\IZ{\Bbb{Z}}
\def\IC{\Bbb{C}}
\def\IR{\Bbb{R}}
\def\IH{\Bbb{H}}
\def\C{{\cal C}}
\def\cP{{\cal P}}
\def\cM{{\cal M}}
\def\cG{{\cal G}}
\def\G{\Gamma}
\def\g{\gamma}
\def\cF{{\cal F}}

\def\Fb{\overline{F}}

\def\cR{{\cal R}}
\def\N{{\cal N}}
\def\la{\langle}
\def\ra{\rangle}

\def\hsl{hs[$\lambda$]}

\def\hsh{hs$\left[{1\over 2}\right]$}

\def\p{\partial}
\def\Wc{\cal{W}}

\def\Oc{{\cal O}}
\def\half{{1\over 2}}
\def\rar{\rightarrow}

\def\a{\alpha}
\def\b{\beta}
\def\o{\omega}
\def\l{\lambda}
\def\t{\tau}
\def\eps{\epsilon}

\def\Ab{\overline{A}}

\def\zb{\overline{z}}
\def\taub{\overline{\tau}}

\def\vs{\vskip .1 in}
\def\bul{$\bullet$~}

\def\IZ{\Bbb{Z}}



\newcount\figno
\figno=0
\def\fig#1#2#3{
\par\begingroup\parindent=0pt\leftskip=1cm\rightskip=1cm\parindent=0pt
\baselineskip=11pt
\global\advance\figno by 1
\midinsert
\epsfxsize=#3
\centerline{\epsfbox{#2}}
\vskip -21pt
{\bf Fig.\ \the\figno: } #1\par
\endinsert\endgroup\par
}
\def\figlabel#1{\xdef#1{\the\figno}}
\def\encadremath#1{\vbox{\hrule\hbox{\vrule\kern8pt\vbox{\kern8pt
\hbox{$\displaystyle #1$}\kern8pt}
\kern8pt\vrule}\hrule}}

\def\p{\partial}

\def\Oc{{\cal O}}

\def\zb{\overline{z}}

\def\psib{\overline{\psi}}

\def\taub{\overline{\tau}}

\def\eps{\epsilon}

\def\Ab{\overline{A}}
\def\Fb{\overline{F}}
\def\zb{\overline{z}}

\def\Wc{{\cal W}}

\def\hb{\overline{h}}

\def\IZ{\Bbb{Z}}
\def\IC{\Bbb{C}}
\def\IR{\Bbb{R}}

\lref\BowcockWQ{
  P.~Bowcock and G.~M.~T.~Watts,
  ``Null vectors, three point and four point functions in conformal field theory,''
Theor.\ Math.\ Phys.\  {\bf 98}, 350 (1994), [Teor.\ Mat.\ Fiz.\  {\bf 98}, 500 (1994)].
[hep-th/9309146].
}

\lref\PiatekIFA{
  M.~Piatek,
  ``Classical torus conformal block, N=2* twisted superpotential and the accessory parameter of Lame equation,''
[arXiv:1309.7672 [hep-th]].
}

\lref\BeccariaLQA{
  M.~Beccaria and G.~Macorini,
  ``On the next-to-leading holographic entanglement entropy in $AdS_{3}/CFT_{2}$,''
[arXiv:1402.0659 [hep-th]].
}

\lref\BagchiVR{
  A.~Bagchi, S.~Lal, A.~Saha and B.~Sahoo,
  ``Topologically Massive Higher Spin Gravity,''
JHEP {\bf 1110}, 150 (2011).
[arXiv:1107.0915 [hep-th]].
}

\lref\ZamolodchikovIE{
  A.~B.~Zamolodchikov,
  ``Conformal Symmetry In Two-dimensions: An Explicit Recurrence Formula For The Conformal Partial Wave Amplitude,''
Commun.\ Math.\ Phys.\  {\bf 96}, 419 (1984)..
}

\lref\CalabreseQY{
  P.~Calabrese and J.~Cardy,
  ``Entanglement entropy and conformal field theory,''
J.\ Phys.\ A {\bf 42}, 504005 (2009).
[arXiv:0905.4013 [cond-mat.stat-mech]].
}
\lref\SolodukhinAH{
  S.~N.~Solodukhin,
  ``Holography with gravitational Chern-Simons,''
Phys.\ Rev.\ D {\bf 74}, 024015 (2006).
[hep-th/0509148].
}

\lref\SunUF{
  J.~-R.~Sun,
  ``Note on Chern-Simons Term Correction to Holographic Entanglement Entropy,''
JHEP {\bf 0905}, 061 (2009).
[arXiv:0810.0967 [hep-th]].
}

\lref\Figueroa{
  J.~M.~Figueroa-O'Farrill, J.~Mas and E.~Ramos,
  ``A One parameter family of Hamiltonian structures for the KP hierarchy and a continuous deformation of the nonlinear W(KP) algebra,''
Commun.\ Math.\ Phys.\  {\bf 158}, 17 (1993).
[hep-th/9207092].
}

\lref\HungNU{
  L.~-Y.~Hung, R.~C.~Myers, M.~Smolkin and A.~Yale,
  ``Holographic Calculations of Renyi Entropy,''
JHEP {\bf 1112}, 047 (2011).
[arXiv:1110.1084 [hep-th]].
}
\lref\HartmanMIA{
  T.~Hartman,
  ``Entanglement Entropy at Large Central Charge,''
[arXiv:1303.6955 [hep-th]].
}

\lref\BarrellaWJA{
  T.~Barrella, X.~Dong, S.~A.~Hartnoll and V.~L.~Martin,
  ``Holographic entanglement beyond classical gravity,''
[arXiv:1306.4682 [hep-th]].
}

\lref\SahooVZ{
  B.~Sahoo and A.~Sen,
  ``BTZ black hole with Chern-Simons and higher derivative terms,''
JHEP {\bf 0607}, 008 (2006).
[hep-th/0601228].
}

\lref\TachikawaSZ{
  Y.~Tachikawa,
  ``Black hole entropy in the presence of Chern-Simons terms,''
Class.\ Quant.\ Grav.\  {\bf 24}, 737 (2007).
[hep-th/0611141].
}

\lref\KrausZM{
  P.~Kraus and F.~Larsen,
  ``Holographic gravitational anomalies,''
JHEP {\bf 0601}, 022 (2006).
[hep-th/0508218].
}

\lref\CarlipQH{
  S.~Carlip,
  ``The Constraint Algebra of Topologically Massive AdS Gravity,''
JHEP {\bf 0810}, 078 (2008).
[arXiv:0807.4152 [hep-th]].
}

\lref\FateevAB{
  V.~A.~Fateev and A.~V.~Litvinov,
  ``Correlation functions in conformal Toda field theory. I.,''
JHEP {\bf 0711}, 002 (2007).
[arXiv:0709.3806 [hep-th]].
}

\lref\FaulknerYIA{
  T.~Faulkner,
  ``The Entanglement Renyi Entropies of Disjoint Intervals in AdS/CFT,''
[arXiv:1303.7221 [hep-th]].
}

\lref\FateevGS{
  V.~A.~Fateev and A.~V.~Litvinov,
  ``On differential equation on four-point correlation function in the Conformal Toda Field Theory,''
JETP Lett.\  {\bf 81}, 594 (2005), [Pisma Zh.\ Eksp.\ Teor.\ Fiz.\  {\bf 81}, 728 (2005)].
[hep-th/0505120].
}

\lref\ChenVP{
  B.~Chen, J.~Long and J.~-B.~Wu,
  ``Spin-3 Topological Massive Gravity,''
Phys.\ Lett.\ B {\bf 705}, 513 (2011).
[arXiv:1106.5141 [hep-th]].
}

\lref\GiombiVD{
  S.~Giombi, A.~Maloney and X.~Yin,
  ``One-loop Partition Functions of 3D Gravity,''
JHEP {\bf 0808}, 007 (2008).
[arXiv:0804.1773 [hep-th]].
}

\lref\YinGV{
  X.~Yin,
  ``Partition Functions of Three-Dimensional Pure Gravity,''
Commun.\ Num.\ Theor.\ Phys.\  {\bf 2}, 285 (2008).
[arXiv:0710.2129 [hep-th]].
}
\lref\GaberdielVVA{
  M.~R.~Gaberdiel and R.~Gopakumar,
  ``Large $\N=4$ Holography,''
JHEP {\bf 1309}, 036 (2013).
[arXiv:1305.4181 [hep-th]].
}

\lref\GaberdielUJ{
  M.~R.~Gaberdiel and R.~Gopakumar,
  ``Minimal Model Holography,''
J.\ Phys.\ A {\bf 46}, 214002 (2013).
[arXiv:1207.6697 [hep-th]].
}

\lref\GaberdielKU{
  M.~R.~Gaberdiel and R.~Gopakumar,
  ``Triality in Minimal Model Holography,''
JHEP {\bf 1207}, 127 (2012).
[arXiv:1205.2472 [hep-th]].
}

\lref\GaberdielWB{
  M.~R.~Gaberdiel and T.~Hartman,
  ``Symmetries of Holographic Minimal Models,''
JHEP {\bf 1105}, 031 (2011).
[arXiv:1101.2910 [hep-th]].
}

\lref\GaberdielPZ{
  M.~R.~Gaberdiel and R.~Gopakumar,
  ``An AdS$_3$ Dual for Minimal Model CFTs,''
Phys.\ Rev.\ D {\bf 83}, 066007 (2011).
[arXiv:1011.2986 [hep-th]].
}

\lref\CastroKE{
  A.~Castro, N.~Lashkari and A.~Maloney,
  ``Quantum Topologically Massive Gravity in de Sitter Space,''
JHEP {\bf 1108}, 040 (2011).
[arXiv:1105.4733 [hep-th]].
}

\lref\CastroCE{
  A.~Castro, A.~Lepage-Jutier and A.~Maloney,
  ``Higher Spin Theories in AdS$_3$ and a Gravitational Exclusion Principle,''
JHEP {\bf 1101}, 142 (2011).
[arXiv:1012.0598 [hep-th]].
}
\lref\PapadodimasPF{
  K.~Papadodimas and S.~Raju,
  ``Correlation Functions in Holographic Minimal Models,''
Nucl.\ Phys.\ B {\bf 856}, 607 (2012).
[arXiv:1108.3077 [hep-th]].
}
\lref\GaberdielZW{
  M.~R.~Gaberdiel, R.~Gopakumar, T.~Hartman and S.~Raju,
  ``Partition Functions of Holographic Minimal Models,''
JHEP {\bf 1108}, 077 (2011).
[arXiv:1106.1897 [hep-th]].
}

\lref\FateevQA{
  V.~Fateev and S.~Ribault,
  ``The Large central charge limit of conformal blocks,''
JHEP {\bf 1202}, 001 (2012).
[arXiv:1109.6764 [hep-th]].
}

\lref\ProkushkinBQ{
  S.~F.~Prokushkin and M.~A.~Vasiliev,
  ``Higher spin gauge interactions for massive matter fields in 3-D AdS space-time,''
Nucl.\ Phys.\ B {\bf 545}, 385 (1999).
[hep-th/9806236].
}
\lref\KrausDS{
  P.~Kraus and E.~Perlmutter,
  ``Partition functions of higher spin black holes and their CFT duals,''
JHEP {\bf 1111}, 061 (2011).
[arXiv:1108.2567 [hep-th]].
}
\lref\DavidXG{
  J.~R.~David, M.~RGaberdiel and R.~Gopakumar,
  ``The Heat Kernel on AdS(3) and its Applications,''
JHEP {\bf 1004}, 125 (2010).
[arXiv:0911.5085 [hep-th]].
}

\lref\KrausVZ{
  P.~Kraus and F.~Larsen,
  ``Microscopic black hole entropy in theories with higher derivatives,''
JHEP {\bf 0509}, 034 (2005).
[hep-th/0506176].
}

\lref\SaidaEC{
  H.~Saida and J.~Soda,
  ``Statistical entropy of BTZ black hole in higher curvature gravity,''
Phys.\ Lett.\ B {\bf 471}, 358 (2000).
[gr-qc/9909061].
}

\lref\PerlmutterDS{
  E.~Perlmutter, T.~Prochazka and J.~Raeymaekers,
  ``The semiclassical limit of W$_N$ CFTs and Vasiliev theory,''
JHEP {\bf 1305}, 007 (2013).
[arXiv:1210.8452 [hep-th]].
}
\lref\ChangIZP{
  C.~-M.~Chang and X.~Yin,
  ``A semi-local holographic minimal model,''
Phys.\ Rev.\ D {\bf 88}, 106002 (2013).
[arXiv:1302.4420 [hep-th]].
}
\lref\ChangMZ{
  C.~-M.~Chang and X.~Yin,
  ``Higher Spin Gravity with Matter in AdS$_3$ and Its CFT Dual,''
JHEP {\bf 1210}, 024 (2012).
[arXiv:1106.2580 [hep-th]].
}
\lref\GaberdielYB{
  M.~R.~Gaberdiel, T.~Hartman and K.~Jin,
  ``Higher Spin Black Holes from CFT,''
JHEP {\bf 1204}, 103 (2012).
[arXiv:1203.0015 [hep-th]].
}

\lref\VasilievBA{
  M.~A.~Vasiliev,
  ``Higher spin gauge theories: Star product and AdS space,''
In *Shifman, M.A. (ed.): The many faces of the superworld* 533-610.
[hep-th/9910096].
}
\lref\GukovYM{
  S.~Gukov, E.~Martinec, G.~W.~Moore and A.~Strominger,
  ``The Search for a holographic dual to AdS(3) x S**3 x S**3 x S**1,''
Adv.\ Theor.\ Math.\ Phys.\  {\bf 9}, 435 (2005).
[hep-th/0403090].
}

\lref\AnninosNJA{
  D.~Anninos, J.~Samani and E.~Shaghoulian,
  ``Warped Entanglement Entropy,''
[arXiv:1309.2579 [hep-th]].
}
\lref\ChenDXA{
  B.~Chen, J.~Long and J.~-j.~Zhang,
  ``Holographic R\'enyi entropy for CFT with $W$ symmetry,''
[arXiv:1312.5510 [hep-th]].
}

\lref\CampoleoniHG{
  A.~Campoleoni, S.~Fredenhagen and S.~Pfenninger,
  ``Asymptotic W-symmetries in three-dimensional higher-spin gauge theories,''
JHEP {\bf 1109}, 113 (2011).
[arXiv:1107.0290 [hep-th]].
}

\lref\BanerjeeAJ{
  S.~Banerjee, A.~Castro, S.~Hellerman, E.~Hijano, A.~Lepage-Jutier, A.~Maloney and S.~Shenker,
  ``Smoothed Transitions in Higher Spin AdS Gravity,''
Class.\ Quant.\ Grav.\  {\bf 30}, 104001 (2013).
[arXiv:1209.5396 [hep-th]].
}
\lref\MaldacenaBW{
  J.~M.~Maldacena and A.~Strominger,
  ``AdS(3) black holes and a stringy exclusion principle,''
JHEP {\bf 9812}, 005 (1998).
[hep-th/9804085].
}

\lref\CampoleoniHP{
  A.~Campoleoni, S.~Fredenhagen, S.~Pfenninger and S.~Theisen,
  ``Towards metric-like higher-spin gauge theories in three dimensions,''
J.\ Phys.\ A {\bf 46}, 214017 (2013).
[arXiv:1208.1851 [hep-th]].
}
\lref\CampoleoniZQ{
  A.~Campoleoni, S.~Fredenhagen, S.~Pfenninger and S.~Theisen,
  ``Asymptotic symmetries of three-dimensional gravity coupled to higher-spin fields,''
JHEP {\bf 1011}, 007 (2010).
[arXiv:1008.4744 [hep-th]].
}
\lref\GaberdielCCA{
  M.~R.~Gaberdiel, R.~Gopakumar and M.~Rangamani,
  ``The Spectrum of Light States in Large N Minimal Models,''
[arXiv:1310.1744 [hep-th]].
}

\lref\HeadrickFK{
  M.~Headrick, A.~Lawrence and M.~Roberts,
  ``Bose-Fermi duality and entanglement entropies,''
J.\ Stat.\ Mech.\  {\bf 1302}, P02022 (2013).
[arXiv:1209.2428 [hep-th]].
}

\lref\LiDQ{
  W.~Li, W.~Song and A.~Strominger,
  ``Chiral Gravity in Three Dimensions,''
JHEP {\bf 0804}, 082 (2008).
[arXiv:0801.4566 [hep-th]].
}

\lref\RajabpourPT{
  M.~A.~Rajabpour and F.~Gliozzi,
  ``Entanglement Entropy of Two Disjoint Intervals from Fusion Algebra of Twist Fields,''
J.\ Stat.\ Mech.\  {\bf 1202}, P02016 (2012).
[arXiv:1112.1225 [hep-th]].
}

\lref\HaydenAG{
  P.~Hayden, M.~Headrick and A.~Maloney,
  ``Holographic Mutual Information is Monogamous,''
Phys.\ Rev.\ D {\bf 87}, no. 4, 046003 (2013).
[arXiv:1107.2940 [hep-th]].
}

\lref\HeadrickZT{
  M.~Headrick,
  ``Entanglement Renyi entropies in holographic theories,''
Phys.\ Rev.\ D {\bf 82}, 126010 (2010).
[arXiv:1006.0047 [hep-th]].
}

\lref\KrausWN{
  P.~Kraus,
  ``Lectures on black holes and the AdS(3) / CFT(2) correspondence,''
Lect.\ Notes Phys.\  {\bf 755}, 193 (2008).
[hep-th/0609074].
}

\lref\BalasubramanianSN{
  V.~Balasubramanian, P.~Kraus and A.~E.~Lawrence,
  ``Bulk versus boundary dynamics in anti-de Sitter space-time,''
Phys.\ Rev.\ D {\bf 59}, 046003 (1999).
[hep-th/9805171].
}

\lref\CalabreseQY{
  P.~Calabrese and J.~Cardy,
  ``Entanglement entropy and conformal field theory,''
J.\ Phys.\ A {\bf 42}, 504005 (2009).
[arXiv:0905.4013 [cond-mat.stat-mech]].
}
\lref\BouwknegtWG{
  P.~Bouwknegt and K.~Schoutens,
  ``W symmetry in conformal field theory,''
Phys.\ Rept.\  {\bf 223}, 183 (1993).
[hep-th/9210010].
}

\lref\HeadrickZT{
  M.~Headrick,
  ``Entanglement Renyi entropies in holographic theories,''
Phys.\ Rev.\ D {\bf 82}, 126010 (2010).
[arXiv:1006.0047 [hep-th]].
}

\lref\RyuBV{
  S.~Ryu and T.~Takayanagi,
  ``Holographic derivation of entanglement entropy from AdS/CFT,''
Phys.\ Rev.\ Lett.\  {\bf 96}, 181602 (2006).
[hep-th/0603001].
}

\lref\LewkowyczNQA{
  A.~Lewkowycz and J.~Maldacena,
  ``Generalized gravitational entropy,''
[arXiv:1304.4926 [hep-th]].
}

\lref\NishiokaUN{
  T.~Nishioka, S.~Ryu and T.~Takayanagi,
  ``Holographic Entanglement Entropy: An Overview,''
J.\ Phys.\ A {\bf 42}, 504008 (2009).
[arXiv:0905.0932 [hep-th]].
}

\lref\VanRaamsdonkAR{
  M.~Van Raamsdonk,
  ``Comments on quantum gravity and entanglement,''
[arXiv:0907.2939 [hep-th]].
}

\lref\CardyIE{
  J.~L.~Cardy,
  ``Operator Content of Two-Dimensional Conformally Invariant Theories,''
Nucl.\ Phys.\ B {\bf 270}, 186 (1986)..
}

\lref\AlvarezGaumeIG{
  L.~Alvarez-Gaume and E.~Witten,
  ``Gravitational Anomalies,''
Nucl.\ Phys.\ B {\bf 234}, 269 (1984)..
}

\lref\MaloneyCK{
  A.~Maloney, W.~Song and A.~Strominger,
  ``Chiral Gravity, Log Gravity and Extremal CFT,''
Phys.\ Rev.\ D {\bf 81}, 064007 (2010).
[arXiv:0903.4573 [hep-th]].
}
\lref\HenneauxXG{
  M.~Henneaux and S.~-J.~Rey,
  ``Nonlinear $W_{infinity}$ as Asymptotic Symmetry of Three-Dimensional Higher Spin Anti-de Sitter Gravity,''
JHEP {\bf 1012}, 007 (2010).
[arXiv:1008.4579 [hep-th]].
}
\lref\HolzheyWE{
  C.~Holzhey, F.~Larsen and F.~Wilczek,
  ``Geometric and renormalized entropy in conformal field theory,''
Nucl.\ Phys.\ B {\bf 424}, 443 (1994).
[hep-th/9403108].
}
\lref\DeserWH{
  S.~Deser, R.~Jackiw and S.~Templeton,
  ``Topologically Massive Gauge Theories,''
Annals Phys.\  {\bf 140}, 372 (1982), [Erratum-ibid.\  {\bf 185}, 406 (1988)], [Annals Phys.\  {\bf 185}, 406 (1988)], [Annals Phys.\  {\bf 281}, 409 (2000)]..
}

\lref\DeserVY{
  S.~Deser, R.~Jackiw and S.~Templeton,
  ``Three-Dimensional Massive Gauge Theories,''
Phys.\ Rev.\ Lett.\  {\bf 48}, 975 (1982)..
}

\lref\JJ{A.~Castro, J. de Boer, J.~Jottar, work in progress}

\lref\tmgp{M.~Ammon, A.~Castro, S.~Detournay, N.~Iqbal, E.~Perlmutter, work in progress}

\lref\CalabreseHE{
  P.~Calabrese, J.~Cardy and E.~Tonni,
  ``Entanglement entropy of two disjoint intervals in conformal field theory II,''
J.\ Stat.\ Mech.\  {\bf 1101}, P01021 (2011).
[arXiv:1011.5482 [hep-th]].
}

\lref\CalabreseEZ{
  P.~Calabrese, J.~Cardy and E.~Tonni,
  ``Entanglement entropy of two disjoint intervals in conformal field theory,''
J.\ Stat.\ Mech.\  {\bf 0911}, P11001 (2009).
[arXiv:0905.2069 [hep-th]].
}

\lref\ElShowkAG{
  S.~El-Showk and K.~Papadodimas,
  ``Emergent Spacetime and Holographic CFTs,''
JHEP {\bf 1210}, 106 (2012).
[arXiv:1101.4163 [hep-th]].
}

\lref\HeemskerkPN{
  I.~Heemskerk, J.~Penedones, J.~Polchinski and J.~Sully,
  ``Holography from Conformal Field Theory,''
JHEP {\bf 0910}, 079 (2009).
[arXiv:0907.0151 [hep-th]].
}

\lref\HarlowNY{
  D.~Harlow, J.~Maltz and E.~Witten,
  ``Analytic Continuation of Liouville Theory,''
JHEP {\bf 1112}, 071 (2011).
[arXiv:1108.4417 [hep-th]].
}



\lref\GaberdielAR{
  M.~R.~Gaberdiel, R.~Gopakumar and A.~Saha,
  ``Quantum $W$-symmetry in $AdS_3$,''
JHEP {\bf 1102}, 004 (2011).
[arXiv:1009.6087 [hep-th]].
}

\lref\HubenyXT{
  V.~E.~Hubeny, M.~Rangamani and T.~Takayanagi,
  ``A Covariant holographic entanglement entropy proposal,''
JHEP {\bf 0707}, 062 (2007).
[arXiv:0705.0016 [hep-th]].
}

\lref\AmmonHBA{
  M.~Ammon, A.~Castro and N.~Iqbal,
  ``Wilson Lines and Entanglement Entropy in Higher Spin Gravity,''
JHEP {\bf 1310}, 110 (2013).
[arXiv:1306.4338 [hep-th]].
}

\lref\deBoerVCA{
  J.~de Boer and J.~I.~Jottar,
  ``Entanglement Entropy and Higher Spin Holography in AdS$_3$,''
[arXiv:1306.4347 [hep-th]].
}

\lref\AnninosFX{
  D.~Anninos, W.~Li, M.~Padi, W.~Song and A.~Strominger,
  ``Warped AdS(3) Black Holes,''
JHEP {\bf 0903}, 130 (2009).
[arXiv:0807.3040 [hep-th]].
}

\lref\zamo{A.~Zamolodchikov, ``Conformal Symmetry in Two-dimensional Space: Recursion Representation of the Conformal Block,'' Teoreticheskaya i
  Matematicheskaya Fizika, {\bf 73, No. 1} (1987).}

\lref\DetournayPC{
  S.~Detournay, T.~Hartman and D.~M.~Hofman,
  ``Warped Conformal Field Theory,''
Phys.\ Rev.\ D {\bf 86}, 124018 (2012).
[arXiv:1210.0539 [hep-th]].
}
\lref\deBoerGZ{
  J.~de Boer and J.~I.~Jottar,
  ``Thermodynamics of Higher Spin Black Holes in AdS$_{3}$,''
[arXiv:1302.0816 [hep-th]].
}
\lref\CalabreseEU{
  P.~Calabrese and J.~L.~Cardy,
  ``Entanglement entropy and quantum field theory,''
J.\ Stat.\ Mech.\  {\bf 0406}, P06002 (2004).
[hep-th/0405152].
}
\lref\BanadosUE{
  M.~Banados, R.~Canto and S.~Theisen,
  ``The Action for higher spin black holes in three dimensions,''
JHEP {\bf 1207}, 147 (2012).
[arXiv:1204.5105 [hep-th]].
}

\lref\DattaHBA{
  S.~Datta and J.~R.~David,
  ``Renyi entropies of free bosons on the torus and holography,''
[arXiv:1311.1218 [hep-th]].
}

\lref\ChenKPA{
  B.~Chen and J.~-J.~Zhang,
  ``On short interval expansion of Rényi entropy,''
JHEP {\bf 1311}, 164 (2013).
[arXiv:1309.5453 [hep-th]].
}



\Title{\vbox{\baselineskip14pt
}} {\vbox{\centerline {Comments on R\'enyi entropy in AdS$_3$/CFT$_2$}}}
\centerline{Eric Perlmutter\foot{E.Perlmutter@damtp.cam.ac.uk}}
\bigskip
\centerline{\it{DAMTP, Centre for Mathematical Sciences,  University of Cambridge}}
\centerline{${}$\it{Cambridge, CB3 0WA, UK}}

\baselineskip14pt

\vskip .3in

\centerline{\bf Abstract}
\vskip.2cm
We extend and refine recent results on R\'enyi entropy in two-dimensional conformal field theories at large central charge. To do so, we examine the effects of higher spin symmetry and of allowing unequal left and right central charges, at leading and subleading order in large total central charge. The result is a straightforward generalization of previously derived formulae, supported by both gravity and CFT arguments. The preceding statements pertain to CFTs in the ground state, or on a circle at unequal left- and right-moving temperatures. For the case of two short intervals in a CFT ground state, we derive certain universal contributions to R\'enyi and entanglement entropy from Virasoro primaries of arbitrary scaling weights, to leading and next-to-leading order in the interval size; this result applies to any CFT. When these primaries are higher spin currents, such terms are placed in one-to-one correspondence with terms in the bulk 1-loop determinants for higher spin gauge fields propagating on handlebody geometries. 

\Date{}
\baselineskip14pt

\listtoc\writetoc

\newsec{Introduction}
Two-dimensional conformal field theories are among the most well-understood quantum field theories. Likewise, their occasional duality with theories of gravity in three-dimensional anti-de Sitter space subjects the AdS/CFT correspondence to some of its most stringent and detailed checks. Nevertheless, we continue to discover new features of these theories, even in their ground states or in holographic limits. If we could understand all physics of the ground state of 2d CFTs at large central charge, this would constitute a major advancement. 

There has been a recent surge of interest in entanglement entropy and the associated R\'enyi entropy, partly due to its promise in helping us achieve that goal. These are examples of quantities that contain all data about a given CFT, processed in a way that emphasizes the way information is organized. Even in the ground state of a CFT, R\'enyi entropies are highly nontrivial; their study provides a quantum entanglement alternative to the conformal bootstrap. 

We can define the ground state R\'enyi entropy, in any spacetime dimension, as follows. At some fixed time, we split space into two subspaces. Upon forming a reduced density matrix $\rho$ by tracing over one subspace, one defines the R\'enyi entropy $S_n$ as
\eqn\dtg{S_n = {1\over 1-n}\log\Tr\rho^n}
where $n\neq 1$ is a positive, real parameter. In the limit $n\rar 1$, this reduces to the von Neumann entropy, otherwise known as the entanglement entropy, $S_{EE}\equiv \lim_{n\rar 1}S_n$. In two dimensions, these spatial regions are unions of $N$ disjoint intervals, with endpoints $z_i$, where $i=1,\ldots,2N$. The R\'enyi entropy can be computed either as the partition function on a higher genus Riemann surface using the replica trick, or as a correlation function of $2N$ twist operators $\Phi_{\pm}$ located at the endpoints $z_i$. (See e.g. \CalabreseQY\ for an overview.)

It was convincingly argued in \refs{\HartmanMIA,\FaulknerYIA}, building on \refs{\RyuBV,\HeadrickZT}, that ground state R\'enyi entropies exhibit universal behavior in certain classes of CFTs at large central charge. This is a strong statement, akin to the Cardy formula for the asymptotic growth of states in any 2d CFT. Subject to certain assumptions that we will describe in detail below, local conformal symmetry is powerful enough to completely fix the R\'enyi entropies for any number of disjoint intervals, at leading order in large central charge. Taking the entanglement limit amounts to a rigorous derivation of the Ryu-Takayanagi formula \refs{\RyuBV,\NishiokaUN} for entanglement entropy in the dual 3d gravity theories, as applied to pure AdS$_3$. Similar conclusions apply to CFTs at finite temperature or on a circle.

It is natural to wonder how far one can push these conclusions. The results of \refs{\HartmanMIA,\FaulknerYIA} apply to chirally symmetric states of CFTs with a sparse spectrum at low dimensions, and equally large left and right central charges, $c_L=c_R$. Are there other families of CFTs in which R\'enyi entropy behaves universally? Do we know the gravity duals of such theories? If we do, can we formulate the arguments in gravitational language? Is R\'enyi entropy also universal in states with asymmetry between left- and right-movers? Furthermore, such generalizations would be useful in honing the holographic entanglement dictionary: if we succeed in generalizing, we can understand which features of these CFTs and their duals participate in the universality, and which do not. Inquiries of a similar spirit were investigated in \refs{\HeadrickZT,\HaydenAG,\HeadrickFK}.

The goal of this paper is to show that one can answer each of these questions affirmatively. We will consider CFTs with higher spin symmetry; with $c_L\neq c_R$; and on the torus with unequal left- and right-moving temperatures. Holographically, we will contend with higher spin gravity, topologically massive gravity and rotating BTZ black holes, respectively. The punchline is that at leading order in large total central charge, the R\'enyi entropies change in a rather straightforward way, if at all, and that the subleading corrections are calculable. 

There are various motivations to consider CFTs with higher spin currents. Higher spin holography has provided valuable perspective on what kinds of CFTs can be expected to have gravity duals; to boot, there are consistent theories of 3d higher spins that lie on the gravity side of specific duality proposals \refs{\ProkushkinBQ,\GaberdielPZ,\GaberdielUJ}. These theories are well-known to present unfamiliar puzzles regarding the roles of spacetime geometry, thermodynamics and gauge invariance.  Likewise, with respect to computing R\'enyi entropy, they present an interesting case \refs{\AmmonHBA,\deBoerVCA}. On the one hand, from the Virasoro perspective the currents are not particularly unusual; the original approach of \refs{\HartmanMIA} neither ruled out nor addressed the possible presence of higher spin currents. On the other, the fact that they furnish an extended chiral algebra implies that the problem may be cast in the language of a larger symmetry, and the bulk side of the story is far less understood than ordinary gravity. It turns out that if we inquire about higher spin CFT R\'enyi entropy for multiple intervals in the ground state, the result is independent of the existence of higher spin currents to {\it leading} order in large $c$. On the CFT side, proving this statement boils down to a slightly nontrivial Virasoro primary counting problem; on the higher spin gravity side, this statement follows from showing that the computation is identical to the one in pure gravity, in which the R\'enyi entropy is computed as a partition function on a handlebody manifold.

A general 2d CFT has $c_L\neq c_R$. The universal dynamics of the current sector of such theories at large $c_L+c_R$ is described by topologically massive gravity (TMG) \refs{\DeserWH,\DeserVY}, another fairly exotic theory. This time, we find that at leading order in large $c_L+c_R$, the R\'enyi entropy only reflects the gravitational anomaly \AlvarezGaumeIG\ for chirally asymmetric states, such as a spatially compact CFT with unequal left- and right-moving temperatures. The manner in which it does so is quite clean, is consistent with the Cardy formula \CardyIE, and implies a formula for the on-shell TMG action evaluated on handlebody geometries. The reason for this simple result is that the conformal block decomposition of a 2d CFT correlation function holomorphically factorizes for each individual conformal family. The new contribution can be chalked up to the fact that the twist fields have spin, proportional to $c_L-c_R$.

Let us give a snapshot of these results in equations, deferring full explanation to the body of the text.\foot{ We use the conventions of \HartmanMIA, although we call accessory parameters $\g_i$, as in \BarrellaWJA, rather than $c_i$. } We mainly have in mind CFTs in their ground state (on the plane), at finite temperature or on a circle (on the cylinder), or both (on the torus). At large $(c_L,c_R)$, in a theory with or without higher spin currents, the R\'enyi entropy $S_n$ is, to leading order,
\eqn\itc{{\p S_n\over \p z_i} = {n\over 6(n-1)}(c_L\g_{L,i}+c_R\g_{R,i})}
The $(\g_{L,i},\g_{R,i})$ are so-called accessory parameters, which partly determine the behavior of the stress tensor components $(T(z),\overline{T}(\zb))$, respectively, in the presence of the twist operator insertions. They are determined by the same trivial monodromy problem given in \refs{\HartmanMIA,\FaulknerYIA} -- one on the left, and one on the right. For fixed $z_i$, the choice of trivial monodromy cycles is determined by a minimization condition, just as in \refs{\HartmanMIA,\FaulknerYIA}. The accessory parameters are also related to the Virasoro vacuum blocks at large $(c_L,c_R)$, which we call $f_{0,L}$ and $f_{0,R}$, respectively:
\eqn\itca{{\p f_{0,L}\over \p z_i} = \g_{L,i}~, ~~ {\p f_{0,R}\over \p z_i} = \g_{R,i}}
When the state is chirally symmetric (e.g. in the ground state), $\g_{L,i}=\g_{R,i}$, only the total central charge contributes, and we return to the result of \refs{\HartmanMIA,\FaulknerYIA}. A familiar example of a chirally asymmetric state to which \itc\ applies is the case of a CFT at finite temperature and chemical potential for angular momentum. When only, say, $c_L$ is large, the leading order result is again \itc, but with only the $c_L$ piece.\foot{The analog of \itc\ was only proven in \refs{\HartmanMIA,\FaulknerYIA} to hold for a noncompact CFT in its ground state and, implicitly, for all states related by conformal transformations. The same is true of our proof of \itc. As in \BarrellaWJA, we take the perspective that it holds more generally, e.g. for a CFT on the torus. However, it should be noted that the latter statement has not been proven, either from a CFT or gravitational point of view.}

At leading order in large central charges, \itc\ is the full answer; what about at next-to-leading order? For the higher spin case, we will focus on the R\'enyi entropy of two intervals in the ground state, and explicitly compute corrections at next-to-leading order in large $c=c_L=c_R$. Working in both CFT and gravity, our methodology is as follows.

First, in CFT, we employ the perturbative short interval expansion of \refs{\CalabreseEZ, \HeadrickZT,\CalabreseHE}. We use it to compute certain universal contributions of a pair of holomorphic and anti-holomorphic spin-$s$ currents to R\'enyi and entanglement entropy, working to leading and next-to-leading order in the interval size. These contributions are manifestly of $O(c^0)$. By ``universal,'' we mean that these terms are generated in any CFT that possesses these currents, although in any particular CFT there may be other ``non-universal'' terms that are of comparable order; this possibility depends on the details of the spectrum and OPE data, as we describe in Subsection 4.2.

In gravity,  AdS/CFT combined with the definition of R\'enyi entropy tells us that the 1-loop free energy of higher spin gauge fields on the appropriate handlebody geometries $\cM$ should be proportional to the $O(c^0)$ contribution to the R\'enyi entropy from their dual currents. Because the higher spin currents do not contribute at $O(c)$ as described earlier, this is in fact their leading contribution. Following \BarrellaWJA, we compute these 1-loop determinants in the small interval expansion using known formulas for handlebodies \GiombiVD\ and spin-$s$ gauge fields \DavidXG, and find complete agreement with CFT. That is, if $S^{(s)}_n$ is the contribution to the CFT R\'enyi entropy from the pair of spin-$s$ currents, and $S^{(s)}_n(\cM)$ is the holographic contribution to the R\'enyi entropy obtained from linearized spin-$s$ gauge field fluctuations about $\cM$, then to the order to which we compute,
\eqn\itd{S^{(s)}_n\big|_{x\ll 1}=S^{(s)}_n(\cM)\big|_{\rm 1-loop} ~.}
The variable $x$ parameterizes the interval size.

We emphasize that this equality holds term-by-term: that is, we will associate a given term in the bulk determinant to the contribution of a {\it specific} CFT operator to the R\'enyi entropy. In all, \itd\ is a satisfying check of our proposal that is quite sensitive to the existence, and choice, of the bulk saddle point: upon picking the right one, the only further ingredient is the most basic tenet of the AdS/CFT correspondence, $Z_{\rm AdS}=Z_{\rm CFT}$. 
\vs
Finally, as a prelude to the $O(c^0)$ results just described, we will derive a result applicable to the short interval expansion in {\it any} CFT: for the case of two short intervals in the CFT ground state, we derive the universal contributions to the R\'enyi and entanglement entropies from Virasoro primaries with {\it arbitrary} scaling weights $(h,\hb)$, at leading and next-to-leading order in the interval size. The explicit expressions can be found in equations (4.24) and (4.25) below. If a CFT contains no Virasoro primaries with conformal dimension $\Delta=h+\hb\leq 1$, these expressions receive no ``non-universal'' corrections to the order to which we work. Our result involves a proposed correction to a result in \CalabreseHE\ regarding the role of spinning (not necessarily chiral) primary operators. In particular, our result is independent of the operator spin, $s=h-\hb$.

To those acquainted with calculations involving R\'enyi entropies, higher spin theories, or TMG, these conclusions may not be surprising. Nonetheless, we believe that they, and their implications, are useful. 

The paper is organized as follows. In Section 2 we review the necessary background material. Section 3 is devoted to R\'enyi entropies in higher spin theories at large $c_L=c_R$. Section 4 zooms out and revisits the short interval expansion as applied to generic spinning primary fields; these results are applied to the higher spin arena in Section 5, where we verify \itd. In Section 6, we consider R\'enyi entropy in chirally asymmetric theories with $c_L\neq c_R$ and in chirally asymmetric states, and its implications for TMG. All of these arguments are supported by gravity and CFT calculations. We conclude in Section 7 with a discussion. Three appendices act as the caboose. \vs

The day this work appeared on the arXiv, so did \ChenDXA, which overlaps with Sections 4 and 5 below.

\newsec{Review of recent progress in 2d CFT R\'enyi entropy}
What follows is a review of the impressive progress made in the last year on computing R\'enyi entropy in 2d CFTs at large central charge. In subsequent sections we will need more details of the CFT arguments than of their gravity counterparts; accordingly, our treatment is lopsided. We refer the reader to the original papers \refs{\HartmanMIA,\FaulknerYIA} for further details and to their predecessors \refs{\CalabreseEU,\RyuBV,\HeadrickZT}, among many others. 

\subsec{CFT at large $c$}
We start on the CFT side, following \HartmanMIA\ in some detail. We work with families of CFTs which, basically speaking, are ``holographic.'' Consider a family of unitary, compact, modular invariant CFTs -- call this family $\C$ -- with $c_L=c_R=c$ that admits a large $c$ limit with the following two key properties. First, it obeys cluster decomposition (and hence has a unique vacuum), and correlation functions are smooth in a finite neighborhood of coincident points. This restricts the OPE coefficients of the theory to scale, at most, exponentially in $c$. Second, the theory has a gap, in the sense that the density of states of dimension $\Delta \lesssim O(c)$ grows polynomially with $c$, at most.\foot{Actually, \HartmanMIA\ makes a stronger claim, but this relaxation is allowed, and we will relax even further in the next section. We thank Tom Hartman for valuable discussions on these matters.} Duals to Einstein gravity with a finite number of light fields have an $O(c^0)$ density of states below the gap, but these are a mere corner of the general space of holographic CFTs \refs{\HeemskerkPN,\ElShowkAG}. 

We now consider computing the R\'enyi entropy in some state that admits use of the replica trick, via twist field correlation functions. For concreteness, consider the case of two disjoint spatial intervals in the ground state, whereby
\eqn\cha{\Tr\rho^n = \la \Phi_+(0)\Phi_-(z)\Phi_+(1)\Phi_-(\infty)\ra_{\C^n/\IZ_n}}
where $z=\zb$. (We have used conformal invariance to fix three positions.) These twist fields have dimensions
\eqn\chb{h_{\Phi}=\hb_{\Phi} = {c\over 24}\left(n-{1\over n}\right)}
The calculation is performed in the orbifold theory $\C^n/\IZ_n$, which inherits a Virasoro $\times$ Virasoro symmetry from $\C$, so we can expand this in Virasoro conformal blocks. In the $z\rar 0$ channel, say,
\eqn\inb{\eqalign{&\la \Phi_+(0)\Phi_-(z)\Phi_+(1)\Phi_-(\infty)\ra_{\C^n/\IZ_n} = \sum_p {(C^p)^2} \cF(h_p,h_{\Phi},nc,z)~\overline{\cF}(\hb_p,\hb_{\Phi},nc,z)}}
The sum is over Virasoro primaries in the untwisted sector of $\C^n/\IZ_n$ labeled by $p$, with $\Phi_+\Phi_-$ OPE coefficients $C^p$. The orbifold theory has central charge $nc$. 

Unitarity, compactness, modular invariance and cluster decomposition tell us that this sum is over a consistent, discrete set of positive energy states with a unique vacuum and good behavior as operators collide. To understand the role of our other assumptions, we take the large $c$ limit. This limit is nontrivial if we hold ${h_{\Phi}/ nc}$ and ${h_{p}/ nc}$ fixed. There is strong evidence \refs{\zamo,\HarlowNY} that in this limit, the Virasoro blocks exponentiate:
\eqn\inc{\lim_{c\rar\infty}\cF(h_p,h_{\Phi},nc,z) \approx \exp\left[-{nc\over 6}f\left({h_p\over nc},{h_{\Phi}\over nc},z\right)\right]}
We will say more about $f$ shortly. For now, we note only that for $h_p\lesssim O(nc)$, it is an increasing function of $h_p$ at fixed $z\ll1$.%

In the large $c$ limit, we can approximate the sum over $p$ in \inb\ by an integral. Defining $\delta_p\equiv h_p/nc$ and $\db_p\equiv \hb_p/nc$,
\eqn\ind{\eqalign{&\la \Phi_+(0)\Phi_-(z)\Phi_+(1)\Phi_-(\infty)\ra_{\C^n/\IZ_n} \cr&\approx \int_0^{\infty} d\d_p\int_0^{\infty} d\overline{\d}_p ~C^2(\d_p,\overline{\d}_p,nc)~ \exp\left[-{nc\over 6}\left(f\left(\d_p,{h_{\Phi}\over nc},z\right)+f\left(\overline{\d}_p,{\hb_{\Phi}\over nc},z\right)\right)\right]}}
This is our key equation. The function $C^2(\d_p,\overline{\d}_p,nc)$ is the sum of all squared OPE coefficients of internal primaries with dimensions $(\delta_p,\overline{\d}_p)$. This definition accounts for any degeneracy $d(p)$ of such operators:
\eqn\ine{C^2(\d_p,\overline{\d}_p,nc) = \sum_{i=1}^{d(p)}(C^p_i)^2}
As our definition makes clear, $C^2(\d_p,\overline{\d}_p,nc)$ depends on the central charge. Note, however, that $C^2(0,0,nc) $ is of order one. 

Having massaged the correlator into the form \ind, what can we conclude? This depends strongly on the growth of the measure, $C^2(\d_p,\overline{\d}_p,nc)$. \HartmanMIA\ made the following observations. First, at any fixed $z$ in the neighborhood of $z=0$, the heavy operators with $\d_p+\db_p> O(c)$ are exponentially suppressed. This is required if the identity is to be the dominant contribution for a finite region around $z=0$. Hence the integration in \ind\ is bounded from above by $\delta_p\sim \db_p\sim O(1)$. 

Second, \ind\ is in fact exponentially dominated by the vacuum contribution, $\d_p=\db_p=0$, in a finite region around $z=0$. Denoting the vacuum block as $f_0\left({h_{\Phi}\over nc},z\right) \equiv f\left(0,{h_{\Phi}\over nc},z\right)$, this means that
\eqn\inf{\la \Phi_+(0)\Phi_-(z)\Phi_+(1)\Phi_-(\infty)\ra_{\C^n/\IZ_n} \approx \exp\left[-{nc\over 6}\left(f_0\left({h_{\Phi}\over nc},z\right)+f_0\left({\hb_{\Phi}\over nc},z\right)\right)\right]~,}
times non-exponential contributions in $c$. The reason is that our assumptions about the bounded growth of states, and of the OPE coefficients, imply that $C^2(\d_p,\overline{\d}_p,nc)$ grows at most exponentially in $c$.\foot{We will examine this statement more closely in Section 3.}

First treating the more manageable case that $C^2(\d_p,\overline{\d}_p,nc)$ grows polynomially with $c$ -- as it does for a theory with an Einstein gravity dual -- \ind\ is exponentially dominated by the $\d_p=\db_p=0$ blocks, on account of the fact that $f$ is an increasing function of $\d_p$ for fixed $z$. We can also expand \ind\ in the $t$-channel, whereupon $z\rar 1-z$, with the same conclusions, this time in a region around $z=1$. The most optimistic perspective is that the integrals are always dominated by the vacuum saddle, $\d_p=\db_p=0$. Then \inf\ holds for $0< z < \half$, at which point there is a transition to the $t$-channel. Using \chb, one thus derives the leading order result for the R\'enyi entropy at small $z$:
\eqn\ing{S_n \approx {nc\over 3(n-1)}f_0\left({h_{\Phi}\over nc},z\right)~, ~~ 0< z < \half}
A parallel analysis holds in the $t$-channel, and the final answer for the R\'enyi entropy, to leading order in large $c$, is then
\eqn\ini{S_n \approx {nc\over 3(n-1)}{\rm min}\left\lbrace f_0\left({h_{\Phi}\over nc},z\right),f_0\left({h_{\Phi}\over nc},1-z\right)\right\rbrace}
The physical statement is that only the vacuum and its Virasoro descendants contribute to the R\'enyi entropy. 

In the more extreme scenario that the measure factor in \ind\ grows like $\exp[c\b(\d_p,\db_p)]$ where $\b(\d_p,\db_p)$ is a $c$-independent function, these conclusions still hold on account of cluster decomposition, but in a smaller region around $z=0$ or $z=1$. The precise size of this region depends on the theory. 
\vs
To actually compute the form of vacuum block $f_0$ connects onto the geometric interpretation of R\'enyi entropy. Twist fields aside, the R\'enyi entropy can be defined in terms of a partition function on a singular genus $g$ Riemann surface, $\Sigma$. One can always write any Riemann surface as a quotient of the complex plane, $\Sigma=\IC/\Gamma$, where $\Gamma$ is an order $g$ discrete subgroup of PSL(2,$\IC$), the M\"obius group. $\G$ is called the Schottky group, and the procedure of writing $\Sigma$ as such a quotient is known as Schottky uniformization; more details in this context can be found in \refs{\HartmanMIA,\FaulknerYIA,\BarrellaWJA}. 

If $v$ is the complex coordinate on the cut surface, the equation that uniformizes the surface is
\eqn\inj{\psi''(v)+T(v)\psi(v)=0}
$T(v)$ can be regarded as the expectation value of the stress tensor in the presence of heavy background fields (to wit, the twist fields). Writing the two linearly independent solutions as $\psi_1(v)$ and $\psi_2(v)$, the coordinate $w=\psi_1(v)/\psi_2(v)$ is single-valued on $\Sigma$.

Specifying to the computation of R\'enyi entropy for two intervals in the ground state, the intervals are bounded by endpoints $\lbrace z_1,z_2,z_3,z_4\rbrace$. $\Sigma$ has genus $g=n-1$. In this case $T(v)$ takes the form \zamo\foot{This equation assumes that the uniformization preserves the $\IZ_n$ replica symmetry.}
\eqn\ink{T(v) =\sum_{i=1}^{4}\left({6h_{\Phi}\over c}{1\over (v-z_i)^2}-{\g_i\over v-z_i}\right)}
The parameters $\g_i$ are called accessory parameters, and they are determined by demanding that $\psi(v)$ has trivial monodromy around $g$ contractible cycles of $\Sigma$. The specific monodromy condition on $\psi_{1,2}(v)$ is discussed in detail in \refs{\HartmanMIA,\FaulknerYIA}; we only wish to emphasize that there is a choice of trivial monodromy cycles. These accessory parameters, finally, are related to the Virasoro vacuum block $f_0$ as
\eqn\inl{{\p f_0\over \p z_i }= \g_i}

The upshot is that we can rewrite \ini\ in terms of the accessory parameters as
\eqn\ini{{\p S_n\over \p z} \approx  {nc\over 3(n-1)}{\rm min}_k\g^{(k)}}
where we minimize over the $s$- and $t$-channel monodromies, indexed by $k$. \ini\ generalizes to any number of intervals \HartmanMIA.

\subsec{Gravity at small $G_N$}
On the bulk side, the goal is in principle straightforward. We want to compute the higher genus partition function holographically, by performing the bulk gravitational path integral over metrics asymptotic to $\Sigma$ with appropriate boundary conditions. This was done in \FaulknerYIA. 

All solutions of pure 3d gravity can be written as quotients of AdS$_3$ (in Euclidean signature, $\IH^3$). In particular, solutions of the form $\cM=\IH^3/\G$ asymptote at conformal infinity to quotients of the plane. So among the contributions to the bulk path integral with $\Sigma$ boundary topology are the so-called handlebody solutions $\cM$, which realize the boundary quotient $\Sigma=\IC/\G$. Handlebodies are not the only bulk solutions with $\Sigma$ on the boundary; but, while unproven, \FaulknerYIA\ motivated the proposal that the holographic R\'enyi entropy should be determined by saddle point contributions of handlebodies. Furthermore, \FaulknerYIA\ assumed that the dominant saddles respect the $\IZ_n$ replica symmetry. The semiclassical approximation of the path integral as a sum over saddle points leads us to identify the renormalized, on-shell bulk action, $S_{\rm grav}$, with minus the log of the CFT partition function; this implies the holographic relation
\eqn\rva{S_n = -{1\over 1-n}S_{\rm grav}(\cM)}

To actually evaluate $S_{\rm grav}(\cM)$ is a fairly technical procedure. The result is that it is fixed by the relation
\eqn\rvaa{{\p S_{\rm grav}(\cM)\over \p z_i} \approx{nc\over 3} \g_i}
where the $\g_i$ are the accessory parameters defining the Schottky uniformization of $\Sigma$ as in \ink. That is, $S_{\rm grav}(\cM)$ is proportional to the large $c$ Virasoro vacuum block. From this perspective, the enforcement of trivial monodromy around specific pairs of points on the boundary is equivalent to demanding the smooth contraction of a specific set of $g$ cycles in the bulk interior. The minimization condition that determines the R\'enyi entropy follows from the usual Hawking-Page-type transition among competing saddles of fixed genus. 

Building on \refs{\HartmanMIA,\FaulknerYIA}, \BarrellaWJA\ generalized this construction to the case of a single interval in a CFT of finite size and at finite temperature $T$, under the assumption that the single-interval version of \ini\ still holds on the torus. Among other results, this showed that finite size effects are invisible at $O(c)$.
\subsec{Quantum corrections}
What about beyond $O(c)$? This is where the details of the particular CFT in question become apparent. In the bulk, the idea is straightforward: AdS/CFT instructs us to compute 1-loop determinants on the handlebody geometries, for whatever bulk fields are present. The formula for such determinants is known for a generic handlebody solution \GiombiVD:
\eqn\ola{Z_{\IH^3/\G} = \prod_{\g\in {\cal P}} \left(Z_{\IH^3/\IZ}(q_{\g})\right)^{1/2}}
The $q_{\g}$ are (squared) eigenvalues of a certain set of elements $\g$ belonging to the Schottky group $\G$; we defer more technical explanation to Section 5. The full bulk answer is a product of \ola\ for all linearized excitations, but the graviton contribution is universal. Unless $\cM$ is the solid torus, this expression is generally not 1-loop exact. 

This technology was put to use in \BarrellaWJA\ for the graviton, with physically interesting results. First, \BarrellaWJA\ provided a systematic algorithm for computing the eigenvalues $q_{\g}$ for the case of two intervals in the ground state, in an expansion in short intervals. (``Short'' means parametrically smaller than any other length scale in the problem; in the ground state of a CFT on a line, the only scale is the interval separation.) This revealed, among other things, a nonvanishing mutual information for widely separated intervals, in contrast to the classical result of Ryu-Takayanagi. \BarrellaWJA\ also showed that for a single interval on the torus, finite size effects do appear at 1-loop. Similar calculations were carried out in \DattaHBA.

The CFT side of this story is also straightforward in principle: one must correct the leading saddle point approximation to \ind. Using results of \CalabreseHE, \ChenKPA\ successfully carried this out for ground state, two interval R\'enyi entropy due to the vacuum block alone, i.e. the exchange of the stress tensor and its descendants. This is sufficient to match to the bulk graviton determinant.

\newsec{R\'enyi entropy in higher spin theories} 
We now turn our attention to theories with higher spin symmetry. Before addressing R\'enyi entropy, we recall some basic ingredients.

\subsec{Rudiments of higher spin holography} 

Let us first define what we mean by a higher spin CFT. The theories we consider obey all assumptions laid out in Section 2 and contain, in addition to the stress tensor, extra holomorphic and anti-holomorphic conserved currents $(J^{(s)},\overline{J}^{(s)})$ of integer spin $s>2$. The holomorphic currents have dimension $(s,0)$, and a planar mode expansion
\eqn\hsa{J^{(s)} = \sum_{n\in \IZ}{J^{(s)}_n\over z^{s+n}}}
and similarly for the anti-holomorphic currents. These currents are Virasoro primary fields,
\eqn\hsb{T(z)J^{(s)}(0) \sim {sJ^{(s)}(0)\over z^2}+{\p J^{(s)}(0)\over z}+\ldots}
The full set of currents furnishes an extended conformal algebra (i.e. a $W$-algebra) with central extension. There exists a zoo of $W$-algebras; for the sake of clarity, we will focus exclusively on CFTs with no more than one current at each spin. In particular we will make use of the $W_N$ algebra, which contains one current at each integer spin $2\leq s \leq N$, as well as the $W_{\infty}[\l]$ algebra \Figueroa, which contains an infinite tower of integer spin currents $s\geq 2$ and a free parameter, $\l$. The simplest such algebra beyond Virasoro ($\equiv W_2$) is the $W_3$ algebra, whose commutation relations we provide in Appendix A for handy reference; see \BouwknegtWG\ for an extensive review of $W$-algebras.

It is convenient to present the representation content of a given CFT using characters. For $W_N$, the Verma module character, which we denote $\chi_{h,N}$, is
\eqn\hsca{\chi_{h,N} = q^{h-c/24}\prod_{n=1}^{\infty}{1\over (1-q^n)^{N-1}}\equiv q^{h-c/24}F(q)^{N-1} }
where $h$ is the $L_0$ eigenvalue of the primary field. The function $F(q)$ is simply the generating function for partitions. The CFT vacuum state $|0\ra$ is invariant under a so-called ``wedge algebra,'' which acts as the analog of sl(2,$\IR$) in the Virasoro case. In particular, it is annihilated by modes
\eqn\hsc{J^{(s)}_{-n}|0\ra=0~, ~~ |n|<s}
For the $W_N$ algebra, the wedge algebra is sl(N,$\IR$). Accordingly we must project these states out of the vacuum representation.  After modding out the null states from the vacuum Verma module $\chi_{0,N}$, it is easy to see that the vacuum character of the $W_{N}$ algebra, $\chi_N\equiv \chi_{0,N}/({\rm null~states})$, is
\eqn\hsd{\chi_{N} = q^{-c/24}\prod_{s=2}^N\prod_{n=s}^{\infty}{1\over 1-q^n}}
At certain values of $c$ at which $W_N$ is the chiral algebra of a rational CFT (e.g. the 3-state Potts model with $W_3$ symmetry), the vacuum representation contains even more null states; at generic $c>N-1$, however, all states counted by \hsd\ have positive norm.

For $W_{\infty}[\l]$, the vacuum character, $\chi_{\infty}$, is the $N\rar\infty$ limit of \hsd, and the wedge algebra is \hsl, a higher spin Lie algebra. For all $W$-algebras, the wedge algebras become proper subalgebras only upon taking $c\rar\infty$ due to ($1/c$ suppressed) nonlinearities. (See Appendix A for the case of $W_3$.) 

Given that our CFT is assumed to not have an exponential (in $c$) number density of light Virasoro primaries in a large $c$ limit, the limit theory can be expected to have a classical gravity dual. The AdS/CFT correspondence tells us that each pair of spin-$s$ currents is dual to a spin-$s$ gauge field in AdS$_3$, denoted $\varphi^{(s)}$. Like ordinary AdS gravity, theories of pure higher spins can be efficiently written using two copies of Chern-Simons theory with gauge algebra $\cG\times \cG$,
\eqn\hse{S_{bulk} = {k\over 4\pi}\left(S_{CS}[A]-S_{CS}[\Ab]\right)}
with
\eqn\hsf{S_{CS}[A] = \int \Tr\left(A\wedge dA + {2\over 3}A\wedge A \wedge A\right)}
where $\Tr$ is the invariant quadratic bilinear form of $\cG$. Also like ordinary AdS gravity, these fields are topological and form a consistent (classical) theory by themselves. This theory can also be supersymmetrized, non-Abelianized, and/or consistently coupled to propagating matter. The latter defines the Vasiliev theories of higher spins \refs{\ProkushkinBQ,\VasilievBA}.

The realization of symmetries in this formulation is quite elegant: $\cG\times \cG$ is identified with the wedge algebra of the CFT, and the asymptotic symmetry of the theory with suitably defined AdS boundary conditions is the extension of $\cG\times \cG$ beyond the wedge to the full higher spin conformal algebra \refs{\CampoleoniZQ,\HenneauxXG,\GaberdielWB}. The generators of $\cG$, which we label $V^s_n$, are identified with the wedge modes $J^{(s)}_n$, with $|n|<s$, and the level $k$ is proportional to the CFT central charge $c=3\ell/2G$. The vacuum character \hsd\ arises as (a chiral half of) the AdS$_3$ 1-loop determinant of the sl(N,$\IR)~\times~$sl(N,$\IR$) theory \GaberdielAR. 

One can make contact with a familiar metric-like formulation in terms of tensors as follows. Expanding the connections along the internal directions,
\eqn\hsg{A = \sum_{s}\sum_{|n|<s}A^{(s)}_n(x^{\mu})V^s_n~, ~~ \Ab = \sum_{s}\sum_{|n|<s}\Ab^{(s)}_n(x^{\mu})V^s_n}
and similarly for the barred connection, one associates the spin-$s$ components of the connection with the spin-$s$ field, $\varphi^{(s)}$. One can form a $\cG$-valued vielbein and spin connection,
\eqn\hsh{e={\ell(A-\Ab)\over 2}~, ~~ \o = {A+\Ab\over 2}}
whereupon the metric-like fields $\varphi^{(s)}$ are linear combinations of products of trace invariants, of order $s$ in the vielbein. (The precise map beyond low spins is subject to some ambiguities that have yet to be resolved \CampoleoniHG.)

An important point is that we always demand that our theory contain a metric. Thus we require that $\cG\supset$ sl(2,$\IR$); the latter corresponds to metric degrees of freedom, and the theory \hse\ admits a consistent truncation to the pure gravity sector. This feature is not shared by the 4d Vasiliev theory, which explains why AdS-Schwarzchild is not a solution of the 4d theory, but all solutions of pure 3d gravity are solutions of \hse. In particular, this last statement applies to the handlebody solutions used to compute holographic R\'enyi entropies in pure gravity.

\subsec{Bulk arguments}
We now consider the higher spin holographic computation of ground state R\'enyi entropies for multiple disjoint intervals. Thinking of the R\'enyi entropy as proportional to the free energy of the CFT on a prescribed Riemann surface $\Sigma$, we aim to establish the following three facts:\vs

1. The R\'enyi entropy can be computed using a saddle point approximation to the higher spin gravitational path integral.\vs

2. The relevant saddle points are the same ones as in pure gravity, namely, the handlebody solutions considered in \FaulknerYIA.\vs

3. The free energy itself, and therefore the holographic R\'enyi entropy, is the same as in pure gravity. \vs

We start by recalling what exactly it is that we are trying to compute. As we noted, $\Sigma$ can always be described as a quotient of the plane by a subgroup $\Gamma\in$ PSL(2,$\IC$), the global conformal group. This statement is independent of whether there are higher spin currents in the CFT. We are sitting in the ground state, so there is no higher spin charge.

So our goal is to compute the bulk path integral over field configurations that include a metric which asymptotes to $\Sigma=\IC/\Gamma$ at conformal infinity, and higher spin fields which vanish there. In ordinary gravity, these metrics are the quotients $\cM=\IH^3/\G$, which remain solutions of the higher spin theory. Are there other configurations in the higher spin theory that satisfy the requisite boundary conditions? In particular, the higher spin theory admits many more solutions that are quotients of AdS$_3$, and we need to address whether any of these enters the computation of R\'enyi entropy. %

The null vectors in the CFT vacuum Verma module map to a $\cG\times\cG$ symmetry of AdS$_3$. This bulk symmetry action induces a boundary action of the wedge algebra. The latter includes the subalgebra of global conformal transformations, whose generators are represented as differential bulk operators. But $\cG$ is larger than sl(2,$\IR$), and the remainder of the symmetry cannot be described as pure coordinate transformations. This is a global version of the statement that higher spin gauge symmetries mix diffeomorphisms with other, non-geometric symmetries.\foot{See \FateevQA\ for a discussion of how to think of the action of sl(3,$\IR$) transformations on CFT observables.} Given that the construction of a Riemann surface is purely geometric, none of the $\cG$ symmetries sitting outside of the sl(2,$\IR$) subalgebra is relevant -- these are bona fide higher spin symmetries. Put another way, bulk quotients $\IH^3/\G'$ with $\G'\notin$ sl(2,$\IR$) will turn on higher spin fields and/or charge. Therefore, we only focus on Einstein quotients of $\IH^3$.%

Having localized the problem to the pure gravity sector, it is plain to see the final result. First, we follow \FaulknerYIA\ in assuming that the replica-symmetric handlebodies $\cM$ dominate the path integral. So our goal is to evaluate the on-shell action of the higher spin theory on the handlebodies $\cM$. But this must take the same value as in pure gravity, due to the allowed consistent truncation of the higher spin fields, $\varphi^{(s)}=0$. 

Perhaps it is helpful to write an equation. In \CampoleoniHP, the authors converted the Chern-Simons action \hse\ for $\cG=$~sl(3,$\IR$) to metric-like language, working to quadratic order in the spin-3 field, $\varphi\equiv \varphi^{(3)}$, and to all orders in the metric. Assuming invertibility of the gravitational vielbein, the result is
\eqn\hsi{S = {1\over 16\pi G}\int d^3x \sqrt{-g}\left[(R+2\Lambda)+\varphi\square'\varphi + O(\varphi^4)\right]}
where $\square'$ is a particular two-derivative operator derived in \CampoleoniHP. The pure metric part of the action is simply Einstein-Hilbert, as follows from the existence of a pure gravity subsector. Since the asymptotic chiral $W_3$ algebra has central charge $c=3\ell/2G$, the coefficient of the gravity action is also the same as in pure gravity. It follows that the renormalized on-shell action for the handlebody geometries $\cM$ is the same as in pure gravity.  This action-based argument is not unique to $\cG=$ sl(3,$\IR$).

We emphasize that these conclusions hold even for a bulk theory with an infinite tower of higher spins, such as the $\cG=$ \hsl\ theory.\foot{When these are further coupled to matter a la Vasiliev, the situation is less clear. In particular, the Vasiliev path integral is not expected to admit a clean description in terms of saddle points. In the discussion section, we will remark on this in the context of the proposed duality of 3d Vasiliev theory with $W_N$ minimal models at large $c$.}

In Section 5, we will check this proposal at the quantum level by computing 1-loop determinants of higher spin fields on $\cM$ and showing that they match a dual CFT calculation. Had we chosen the wrong saddle -- or if there was no dominant saddle point at all -- that agreement would disappear.  

\subsec{CFT arguments}
An efficient strategy here is to phrase the problem in Virasoro language, rather than $W$-language.\foot{Organizing the problem using the full $W$ symmetry of the theory turns out to be unproductive, on account of complications of $W$-algebra representation theory. We briefly discuss such an approach in Appendix C.} In a $W$-algebra, the $W$ currents are Virasoro primary (see Appendix A for the $W_3$ algebra, for instance). So are many of their $W$-descendants. If we can show that there are not too many light Virasoro primaries as a function of $c$, then the arguments of \HartmanMIA\ apply without modification. 

A more precise statement is as follows: as long as the growth of light Virasoro primaries at some fixed dimension in the orbifold theory $\C^n/\IZ_n$ is less than $\exp(\b c)$ for a $c$-independent constant $\b$, the measure factor in \ind\ does not grow fast enough to spoil the conclusions of Section 2, and the R\'enyi entropy is still given by the original results obtained in the absence of higher spin symmetry.  The twist field correlators involved in the R\'enyi entropy computations are performed in the orbifold theory $\C^n/\IZ_n$. This theory has many more Virasoro primaries than $\C$, and it is {\it these} states that run in the virtual channels whose growth we need to count. We now show that the above condition is met by CFTs obeying our assumptions that have essentially any $W$-symmetry.

Let us first phrase the problem for a {\it general} $W$-algebra. The spectrum of a CFT $\C$ with (two copies of) a $W$-symmetry can be organized into irreducible representations of the $W$-algebra. Accordingly the partition function can be written as a sum over its highest weight characters. Highest weight representations may contain null states, but these are absent at large $c$ for a generic representation, and we remain indifferent about their possible appearance for now. We thus write a generic partition function as
\eqn\wra{\eqalign{Z_{\C} &= \Tr\left(q^{L_0-c/24}\overline{q}^{\overline{L}_0-c/24}\right) \cr&= |\widehat{\chi}_{0}|^2 + \sum_{h,\hb}\N_{h\hb}\widehat{\chi}_{h}\overline{\widehat{\chi}}_{\hb}}}
The trace is over the entire spectrum of $W$-primaries. $\widehat{\chi}_{0}$ is the vacuum character, and $\widehat{\chi}_h$ is the character of an irreducible highest weight representation labeled by dimension $h$ (and other $W$-quantum numbers that do not feature in this trace). As for the multiplicities $\N_{h\hb}$, we require only that for dimensions populating the $O(c)$ gap, they grow slower than $\exp(\b c)$ as we take $c\rar\infty$, where $\b$ is a $c$-independent constant.   

We are interested in counting light Virasoro primaries with $h\lesssim O(c)$ in the orbifold theory $\C^n/\IZ_n$ at large $c$. We ignore the orbifold for the time being and focus on the tensor product theory $\C^n$, which has partition function $Z_{\C^n}=(Z_{\C})^n$. Using \wra, this can be expanded as a sum over terms of the form
\eqn\ptb{(\widehat{\chi}_{0}\overline{\widehat{\chi}}_{0})^{n-k}\prod_{i=1}^k\N_{h_i\hb_i}\widehat{\chi}_{h_i}\overline{\widehat{\chi}}_{\hb_i}\times {\rm (Constant)}~, ~~ 0\leq k \leq n}
$(h_i,\hb_i)$ can take any allowed values in the spectrum of $\C$. If we continue to ignore the orbifold, we can place a crude upper bound on the growth of light Virasoro primaries by branching \ptb\ into Virasoro highest weight characters, and estimating their growth at large dimension. The constant is a combinatoric factor we will not need, and our assumption about $\N_{h_i\hb_i}$ allows us to ignore that too.

In general, it is a simple matter to branch a given character $\chi$ into generic Virasoro highest weight characters, which are given in \hsca\ with $N=2$,
\eqn\wce{\chi_h \equiv \chi_{h,2}= q^{h-c/24} F(q) }
To execute the branching of $\chi$ without including the Virasoro vacuum character in the decomposition, we write $\chi = \sum_h d(h)\chi_h$. Using \wce, the generating function for Virasoro primaries is then
\eqn\grc{\chi/F(q)=q^{-c/24}\sum_hd(h)q^{h} }
If we wish to include the vacuum in this decomposition,
one simply subtracts $1-q$ from the left-hand side of \grc. Of course, in estimating the growth of $d(h)$ at large $h$, this makes no difference. 

Applying \grc\ for present purposes, we take $\chi$ to be the holomorphic part of \ptb, 
\eqn\ptc{\chi\equiv (\widehat{\chi}_{0})^{n-k}\prod_{i=1}^k\widehat{\chi}_{h_i}~, ~~ 0\leq k \leq n}
and estimate the growth of $d(h)$ at asymptotically large $h$ using saddle point methods. Despite doubly overcounting by computing the growth of Virasoro primaries in the product theory $\C^n$ at asymptotically large $h$ , this will be sufficient in a wide range of cases to show that the growth of Virasoro primaries in $\C^n/\IZ_n$ with $h\lesssim O(c)$ is slow enough to leave the large $c$ R\'enyi entropy unaltered. 
\vs
{\it \bf i) $W_N$ CFT:}\vs
Having presented our algorithm for general higher spin CFTs, we apply it to a CFT with $W_N$ symmetry. The $W_N$ Verma module character ($\chi_{h,N}$) and vacuum character $(\chi_N$) were given in \hsca\ and \hsd, respectively; note that $\chi_N$ can be written conveniently as
\eqn\ptd{\chi_N=q^{-c/24}F(q)^{N-1}\cdot V}
where $V$ is the Vandermonde determinant,
\eqn\pte{V = \prod_{i=2}^N\prod_{j=1}^{i-1}(1-q^{i-j})}
Furthermore, let us assume that the only null states lie in the vacuum representation (this only serves to increase the total number of states), so the highest weight characters are indeed given by $\chi_{h,N}$. Our CFT partition function is then the $W_N$ version of \wra,
\eqn\wrd{Z_{\C} = |\chi_N|^2+\sum_{h,\hb}\N_{h\hb}\chi_{h,N}\overline{\chi}_{\hb,N}}
Consequently, we want to apply our algorithm with the $W_N$ substitutions $\widehat{\chi}_{0} \mapsto \chi_N$ and $\widehat{\chi}_{h_i} \mapsto \chi_{h_i,N}$.

Applying these substitutions to \grc\ and \ptc, we want to estimate the asymptotic growth of $d_{N,n}(h)$ as defined by
\eqn\wre{ q^{x}F(q)^{(N-1)n-1}V^{n-k}=q^{-c/24}\sum_h d_{N,n}(h)q^{h}}
where $x=-nc/24+\sum_{i=1}^kh_i$, and $0\leq k \leq n$. We have labeled the degeneracy $d_{N,n}(h)$ to reflect its dependence on the $n$-fold product and our choice of $W_N$. The $q^x$ factor is not relevant for what follows. 

We see now that we are in the clear: since $V$ is a finite polynomial in $q$, the leading growth is generically controlled by the $F(q)$ factor, which is nothing but a colored partition function with a finite number $(N-1)n-1$ of colors. Colored partitions grow like $d(h)\approx h^{-\a}\exp(2\pi \sqrt{\b h})$ with color-dependent constants $(\a,\b)$. Crucially for us, this growth is slower than $\exp(\b h)$.

A more careful analysis confirms this expectation. A very similar calculation was done in \CastroCE\ using traditional saddle point methods, and we utilize their approach here. The leading order result for the asymptotic growth of $d_{N,n}(h)$ in \wre\ is found to be
\eqn\pth{d_{N,n}(h) \sim h^{-\a}\exp\left(2\pi\sqrt{\b h}\right)~,}
where
\eqn\ptj{\a={3\over 4}+{1\over 4}\left((N^2-1)n+N(N-1)k-1\right)~, ~~ \b = {(N-1)n-1\over 6}}
We are interested in finite $(N,n)$ only, whereupon \pth\ is valid (for $h\gg nN^3$) and the constants $\a$ and $\b$ are finite. Therefore, despite overcounting in several ways, we can conclude that a $W_N$ CFT subject to our assumptions does not lead to an unacceptable number of Virasoro primary states in the orbifold theory $\C^n/\IZ_n$. %

This conclusion extends to any $W$-algebra with a finite number of currents. Actually, we note in passing that this argument is necessary for the original conclusions of \HartmanMIA\ to hold in the Virasoro case, which corresponds to $N=2$ in the above. 
\vs
{\it \bf ii) $W_{\infty}[\l]$ CFT:}\vs
We can even extend this type of argument to the case of $W_{\infty}[\l]$ for generic $\l$. This is slightly trickier because the Verma module has an infinite number of states at all levels above the ground state; some examples of $W_{\infty}[\l]$ highest weight characters of {\it degenerate} representations are given in \refs{\GaberdielZW,\GaberdielKU}. Rather than specifying precisely which of these representations we allow in our CFT, let us just write its partition function as
\eqn\pta{Z_{\C} = |\chi_{\infty}|^2 + \sum_{h,\hb}\N_{h,\hb}\chi_h\chi_{\hb}}
The first term is the vacuum character, given in \hsd\ with $N\rar\infty$. We have implicitly branched all $W_{\infty}[\l]$ highest weight representations of our theory into Virasoro representations, which the second term sums over. In doing so, we only require that $\N_{h\hb}$ does not grow exponentially with $c$ for dimensions populating the gap (which we will further justify in a moment).

Using \pta\ and our previous branching formulas, this time we wish to estimate the growth of 
\eqn\wrp{ q^x F(q)^{k-1}(\chi_{\infty})^{n-k}=q^{-c/24}\sum_h d_{\infty,n}(h)q^{h}}
at large $h$, where $x=-kc/24+\sum_{i=1}^k h_i$. Again using saddle point techniques as in \CastroCE, the asymptotic degeneracy is
\eqn\pth{d_{\infty,n}(h) \sim h^{-\a}\exp\left(h^{2/3}\b\right)~,}
where
\eqn\ptj{\a=\half+{1\over 36}(11k-5n)~, ~~ \b =\left({27\zeta(3)(n-k)\over 4}\right)^{1/3}}
While faster than the $W_N$ growth of primaries, this is not fast enough to spoil the conclusions, despite the infinite tower of spins and our overcounting.\foot{This calculation also partially justifies our assumption about the growth of $\N_{h,\hb}$ in \pta. Characters of a simple class of degenerate representations of $W_{\infty}[\l]$ can be written as $\chi_{\infty}$ times the large $N$ limit of $u(N)$ characters (i.e. characters of \hsl) \refs{\GaberdielZW}. Given our result \pth, these representations do not contain more than $\exp(\b c)$ Virasoro primaries below the gap. This essentially follows from the fact that the \hsl\ characters can be viewed as restricted partitions into a finite set, which have only polynomial asymptotic growth. Thus if the spectrum of $\C$ contains fewer than $\exp(\b c)$ such degenerate representations of $W_{\infty}[\l]$, its branching into Virasoro representations is sufficiently bounded.} 

This same logic applies to all $W$-algebras with an infinite tower of currents, with any finite number of currents at each spin. The latter group includes algebras that should arise as asymptotic symmetries in non-Abelian Vasiliev theories, for example, the $s\tilde{\Wc}_{\infty}^{(4)}[\g]$ algebra with large $\N=4$ supersymmetry that has recently featured in an attempt to connect 3d higher spins to string theory \GaberdielVVA.

\subsec{Looking ahead}
We have argued that the leading order R\'enyi entropy is unaffected by the presence of higher spin currents in the CFT spectrum. These currents do, however, affect the R\'enyi entropy at $O(c^0)$ and beyond, and we would like to compute their contributions and match them to a gravity calculation at 1-loop. Our strategy will be to focus on the case of two intervals in the ground state and perform a short interval expansion. 

\newsec{The short interval expansion for spinning Virasoro primaries}
The short interval expansion of R\'enyi entropies was initialized in \refs{\CalabreseEZ, \HeadrickZT} and presented in full generality by \CalabreseHE. (See also \RajabpourPT.) We will provide a streamlined, and slightly different, version of their presentation here. This will lead us to propose a correction to a result in \CalabreseHE\ regarding the contributions of spinning (not necessarily chiral) Virasoro primary fields. 
\subsec{Generalities} 
Consider a reduced density matrix, $\rho$, obtained by tracing over some degrees of freedom of a CFT we again label $\C$. It was shown that in the short interval expansion, $\Tr\rho^n$ can be written in terms of correlation functions of operators in the spectrum of $\C^n$. Let us quickly establish some notation pertaining to $\C^n$. We refer to operators in the mother CFT, $\C$, as $\Oc$, and to operators of $\C^n$ as $\Och$. If $\C$ has an extended chiral algebra, so does $\C^n$. The modes of the latter, call them $\widehat{W}_n$, are written in terms of the modes of the former, $W_n$, as a sum over $n$ copies of $\C$,
\eqn\chh{\widehat{W}_n = W_n\otimes 1\otimes \ldots \otimes 1 ~+~ {\rm perms}}
Here, 1 denotes the vacuum, and the product is over $n$ sheets of $\C^n$. These modes $\widehat{W}_n$ include the privileged Virasoro modes, $\wh{L}_n$. We will be mostly interested in quasiprimaries of $\C^n$, which obey $\widehat{L}_1\Och=0$. 

Henceforth we specify to the two-interval case in the ground state of $\C$ on the infinite line. As in \cha, we want to compute the orbifold theory twist field four-point function,
\eqn\cha{\Tr\rho^n = \la \Phi_+(z_1)\Phi_-(z_2)\Phi_+(z_3)\Phi_-(z_4)\ra_{\C^n/\IZ_n}}
Using conformal symmetry to fix two of the positions by taking $z_1=0,z_3=1$ and relabeling $z_2=\ell_1,z_4=1+\ell_2$, the cross ratio $x$ is
\eqn\chb{x=\overline{x}={z_{12}z_{34}\over z_{13}z_{24}} = {\ell_1\ell_2\over 1-(\ell_1-\ell_2)}}
where $\ell_{i}$ are the interval lengths obeying $0<\ell_1<1$ and $\ell_2>0$. The essential features of what follows depend only on the cross-ratio, so to make life simpler we take $\ell_1=\ell_2=\ell$; then the small interval expansion is in a small cross-ratio, $x=\ell^2\ll 1$. 

In this regime, the calculation can be efficiently organized as a sum over correlators of quasiprimaries of $\C^n$ and their descendants. This amounts to deriving the fusion rule for the twist operators,
\eqn\chb{[\Phi_+]\times [\Phi_-] = [1] + \sum_K [\Och_K]}
where $K$ indexes quasiprimaries $\lbrace\Och_K\rbrace$ of $\C^n$, and the bracket denotes the entire SO(2,1) $\times$ SO(2,1) conformal family. We take the operators $\Och_K$ to have dimension $\Delta_K$ and norm $\N_K=\la \Och_K|\Och_K\ra$, and we write their OPE coefficients appearing in the fusion \chb\ as $d_K$. Then we can write $\Tr\rho^n$ very neatly as a sum over global conformal blocks:\foot{We have used the fact that $x=\bar x$ to simplify the more general expression for the global conformal blocks first derived in \ZamolodchikovIE. An exchanged field of dimension $\Delta=h+\hb$ and spin $s=h-\hb$ contributes as
\eqn\chba{G_{\Delta,s}(z,\zb) = \half\big|z^h{}_2F_1(h,h;2h;z)\big|^2+(z\leftrightarrow \zb)}
where $z\zb = \left|{z_{12}z_{34}\over z_{13}z_{24}}\right|^2$.}
\eqn\chc{\Tr\rho^n = x^{-2\Delta_{\Phi}}\sum_K {d_K^2\over \N_K}x^{\Delta_K}\big|{}_2F_1(h_K,h_K;2h_K;x)\big|^2}%

The leading term in \chc\ is due to the vacuum exchange, whose entire contribution is simply 1 on account of global conformal symmetry. Pulling this factor out and summing over non-vacuum quasiprimaries only, we can thus write the R\'enyi entropy \dtg\ in the small $x$ expansion as
\eqn\chd{S_n = {c\over 6}\left(1+{1\over n}\right)\log x + {1\over 1-n}\log\left[1+\sum_K {d_K^2\over \N_K}x^{\Delta_K}\big|{}_2F_1(h_K,h_K;2h_K;x)\big|^2\right]}
By Taylor expanding the log, we can isolate the leading contributions of the $K^{'\rm th}$ quasiprimary field to the R\'enyi entropy. In the event that $\Delta_K>1$, the leading and next-to-leading order terms are simply
\eqn\chea{S_n\big|_{\Och_K,\Delta_K>1, x\ll1} \approx{d_K^2\over (1-n)\N_K}x^{\Delta_K}\left(1+{h_K+\hb_K\over 2}x+O(x^2)\right)}
which come from the quasiprimary and its first descendant. We will need this formula later. At any given order in $x$, these terms mix with descendant contributions of other conformal families of lower dimension.

So, computing the contribution of the $K$'th conformal family boils down to finding the OPE coefficients $d_K$ and norm $\N_K$. A key point is that not all quasiprimaries $\Och_K$ have $d_K$ nonzero. This is evident from the formula for the OPE coefficients \CalabreseHE\foot{We use slightly different normalization conventions compared to \ChenKPA\ (they include a factor of the norm, which they call $\a_K$) and \CalabreseHE\ (they orthonormalize their operators).}
\eqn\chf{d_K = \ell^{-\Delta_K}\lim_{z\rar\infty}|z^{2h_K}|^2\la \Och_K(z,\zb)\ra_{\cR_{n,1}}}
The one-point function is computed on the Riemann surface $\cR_{n,1}$, defined as the $n$-sheeted replica surface used to compute the {\it single}-interval R\'enyi entropy for an interval of length $\ell$ on the infinite line. As is well known, this surface can be conformally mapped back to $\IC$ by the transformation
\eqn\chg{z \mapsto f(z) = \left({z-\ell\over z}\right)^{1/n}}
which we can use to compute the one-point functions in \chf. Therefore, $d_K$ is only nonzero if the one-point function of $\Och_K$ on $\cR_{n,1}$ is nonvanishing.

\subsec{Spinning primaries revisited}

In this subsection we address the following question: {\it given the presence of a Virasoro primary field $\Oc$ of scaling dimensions $(h,\hb)$ in the spectrum of $\C$, what are its universal contributions to the R\'enyi entropy to leading and next-to-leading order in the short interval expansion?} By ``universal'' we mean contributions that appear in any CFT containing the operator $\Oc$, irrespective of the remaining spectral and OPE data of that CFT. To answer this, we must enumerate all quasiprimaries $\Och$ that are built out of $\Oc$ with sufficiently low dimension and appear in the $\Phi_+\Phi_-$ OPE. Our answer will also easily admit an $n\rar 1$ limit, which yields spinning primary contributions to  {\it entanglement} entropy.

\vs

{\bf i) Leading order:}\vs

Equation \chf\ implies that the leading contribution from the presence of $\Oc$ in the spectrum of $\C$ comes from a quasiprimary $\Och$ that is made of two copies of $\Oc$ living on different sheets. It is easy to appreciate this point. A general quasiprimary operator $\Och$ of $\C^n$ can be written as a linear combination of operators of the form 
\eqn\chj{\Oc_1(z_1,\zb_1)\otimes \Oc_2(z_2,\zb_2)\otimes\ldots\otimes\Oc_n(z_n,\zb_n)\equiv \prod_{i=1}^n\Oc_i(z_i,\zb_i)}
Because $\la\p^n\Oc\ra_{\cR_{n,1}}\propto \la \p^n\Oc\ra_{\IC}=0$ for all $n$, the first operator of the form \chj\ with nonzero OPE coefficient is made of two $\Oc$ living on different sheets of $\C^n$ \refs{\HeadrickZT,\CalabreseHE}. Let us call this operator $\Och_{\Oc_{j_1}\Oc_{j_2}}$, defined as\foot{To avoid clutter, we have suppressed the sheets on which the operators do not live; these sheets sit in the vacuum, as in \chh.}
\eqn\chk{\Och_{\Oc_{j_1}\Oc_{j_2}} \equiv  \Oc(z_{j_1},\zb_{j_1})\otimes\Oc(z_{j_2},\zb_{j_2})~, ~~ j_1\neq j_2}
There are $n-k$ distinct quasiprimaries $\Och_{\Oc_{j_1}\Oc_{j_2}}$ for which the operators $\Oc$ are spaced by $k$ sheets, $|j_1-j_2|=k$, and the leading order effect on the R\'enyi entropy due to the existence of $\Oc$ is a sum over all such operators. 

First we compute the norm $\N_{\Oc_{j_1}\Oc_{j_2}}$, which is just the square of the normalization of the two point function:
\eqn\chka{\N_{\Oc_{j_1}\Oc_{j_2}}=\la\Oc|\Oc\ra^2}
\chf\ tells us that it will drop out of the R\'enyi entropy \chea, so we need not compute this explicitly. For primary fields, 
\eqn\chl{\la \Oc(z_{j_1},\zb_{j_1})\Oc(z_{j_2},\zb_{j_2}) \ra_{\cR_{n,1}} =\prod_{i=1}^2 f'(z_{j_i})^{h}f'(\zb_{j_i})^{\hb} \cdot \la \Oc(f(z_{j_1}),f(\zb_{j_1}))\Oc(f(z_{j_2}),f(\zb_{j_2})) \ra_{\IC}}
Now, we note that the transformation \chg\ and its derivatives are multi-valued.\foot{This point, also noted in \ChenKPA, explains the discrepancy between what follows and the results of \CalabreseHE.} In the limit $z_{j_i}\rar\infty_{j_i}$, $f(z_{j_i})$ behaves like
\eqn\chm{\eqalign{f(z_{j_i}\rar\infty_{j_i}) &\approx e^{2\pi i j_i/n}\cr
f^{(m)}(z_{j_i}\rar\infty_{j_i}) &\approx e^{2\pi i j_i/n}{\ell m!\over n}(-z_{j_i})^{-m-1}~, ~~ m\neq 0\cr}}
For our operator, \chf\ can be written as
\eqn\chp{\eqalign{d_{\Oc_{j_1}\Oc_{j_2}} =  \left|\prod_{i=1}^2\lim_{z_{j_i}\rar\infty_{j_i}}{z_{j_i}^{2h}\over \ell^h}f'(z_{j_i})^h\right|^2\la \Oc(e^{2\pi i j_1/n})\Oc(e^{2\pi i j_2/n})\ra_{\IC}\cr}}
Combining \chm\ and \chp\ with the form of CFT two-point functions on $\IC$, the OPE coefficient $d_{\Oc_{j_1}\Oc_{j_2}}$ is
\eqn\chq{d_{\Oc_{j_1}\Oc_{j_2}} = \sqrt{\N_{\Oc_{j_1}\Oc_{j_2}}}\left(2n\sin{\pi  j_{12}\over n}\right)^{-2(h+\hb)}}
where $j_{12}\equiv j_1-j_2$. This result is independent of the spin and depends only on the sheet separation. Compared to the same result in \CalabreseHE, it differs by a spin-dependent phase factor. 

With an eye toward our application to higher spin currents, we can consider the case $\Delta(\Och_{\Oc_{j_1}\Oc_{j_2}})=2(h+\hb)>1$. Then plugging \chq\ into \chea, and summing over the full set of distinct operators labeled by $(j_1,j_2)$, we obtain the leading and next-to-leading order contributions to the small $x$ R\'enyi entropy from the quadratic operators $\Och_{\Oc_{j_1}\Oc_{j_2}}$: 
\eqn\chs{S_n\Big|_{\Och_{\Oc\Oc},x\ll1} \approx  {x^{2(h+\hb)}\over 1-n}\sum_{k=1}^{n-1}(n-k)\left({1\over 4n^2\sin^2{\pi k\over n}}\right)^{2(h+\hb)}\left[1+(h+\hb)x+O(x^2)\right]}
where $k=|j_{12}|$. Note that this result is of $O(c^0)$. \vs%

{\bf ii) Next-to-leading order}\vs

The first universal subleading contribution of the operator $\Oc$ comes from the following set of quasiprimaries of dimension $\Delta=2(h+\hb)+1$:
\eqn\qsa{\Och_{\Oc_{j_1}\p\Oc_{j_2}} \equiv \Oc(z_{j_1},\zb_{j_1})\otimes\p\Oc(z_{j_2},\zb_{j_2})-(-1)^{2s}(j_2\leftrightarrow j_1)~, ~~ j_1\neq j_2}
and
\eqn\qsb{\Och_{\Oc_{j_1}\pb\Oc_{j_2}} \equiv  \Oc(z_{j_1},\zb_{j_1})\otimes\pb\Oc(z_{j_2},\zb_{j_2})-(-1)^{2s}(j_2\leftrightarrow j_1)~, ~~ j_1\neq j_2}
where $s=h-\hb$. These are quasiprimaries due to a cancellation between the two terms, and there are again $n-k$ distinct quasiprimaries for which the operators are spaced by $k$ sheets. The $(-1)^{2s}$ factor accounts for possible fermionic statistics. 

There are many possible ``non-universal'' -- that is, theory-specific -- contributions at this or lower order in $x$. For instance, if $\Delta\leq \half$, one could form $\C^n$ quasiprimary operators quartic (or higher) in $\Oc$ that have dimension less than $2\Delta+1$, and which are guaranteed to appear in the twist field fusion rule due to the appearance of the identity in the fusion of two $\Oc$'s. However, the full contribution of such quartic operators would be non-universal, due to the definition of the OPE coefficient \chf\ and the basic fact that four-point functions on the plane are not fixed by conformal symmetry. If $\Delta<1$, an operator of $\C^n$ comprised solely of three $\Oc$'s will contribute to the R\'enyi entropy with a smaller power of $x$ than \qsa\ and \qsb, if the 3-point function coefficient $C_{\Oc\Oc\Oc}$ is nonzero. Similarly, the contributions of \qsa\ and \qsb\ may not be next-to-leading if the CFT possesses other operators {\it besides} $\Oc$ that have conformal dimension less than one. 
In the spirit of universality, we focus on the derivative operators \qsa\ and \qsb\ which  appear in any CFT for any $\Delta$.

Moving on, the treatment of \qsa\ and \qsb\ is quite similar to our leading order analysis, so we will be briefer. We focus on the holomorphic derivative operator \qsa\ only; the analogous contribution from operators \qsb\ is obtained by taking $h\leftrightarrow \hb$ in all that follows. We also assume $(h,\hb)\in \IZ$ for simplicity, but this makes little essential difference. The norm of such states is
\eqn\qsc{\N_{\Och_{\Oc_{j_1}\p\Oc_{j_2}}} = 4h~\N_{\Och_{\Oc_{j_1}\Oc_{j_2}}}}
which one proves using $\widehat{L}_{-1}=\p$ and $\widehat{L}_1\Oc=0$. The OPE coefficient is
\eqn\qsd{d_{\Och_{\Oc_{j_1}\p\Oc_{j_2}}} = 2h{\cos{\pi j_{12}\over n}\over n\sin{\pi j_{12}\over n}}\times \left[\sqrt{\N_{\Oc_{j_1}\Oc_{j_2}}}\left(2n\sin{\pi  j_{12}\over n}\right)^{-2(h+\hb)}\right]}
The factor in brackets is simply the OPE coefficient of the operators $\Och_{\Oc_{j_1}\Oc_{j_2}}$ considered previously, cf. \chq. 

Plugging into \chea, keeping only the leading term and summing over the full set of distinct operators labeled by $(j_1,j_2)$, the leading contribution to the R\'enyi entropy from operators \qsa\ is
\eqn\qse{\eqalign{S_n\Big|_{\Och_{\Oc\p\Oc},x\ll1} \approx~ &{x^{2(h+\hb)}\over 1-n}\sum_{k=1}^{n-1}(n-k)\left({1\over 4n^2\sin^2{\pi k\over n}}\right)^{2(h+\hb)}\left[ xh \left({\cos{\pi k\over n}\over n\sin{\pi k\over n}}\right)^2+O(x^2)\right]}}
Notice that the summand is simply the leading term of \chs, times the factor in brackets.

\subsec{Summary of results and their entanglement entropy limit}

Collecting our results, the {\it universal} R\'enyi entropy contribution $S_n^{(\Delta,s)}$ of a Virasoro primary operator with conformal dimension $\Delta=h+\hb$ and arbitrary spin $s=h-\hb$, to next-to-leading order in small $x$, is given by the sum of \chs, \qse\ and the anti-holomorphic ($h\rar \hb$) version of \qse. Putting these together, we find 
\eqn\sumry{\eqalign{S_n^{(\Delta,s)}\Big|_{x\ll1} \approx  {x^{2\Delta}\over 1-n}\sum_{k=1}^{n-1}(n-k)\left({1\over 4n^2\sin^2{\pi k\over n}}\right)^{2\Delta}\left[1+\Delta x+\Delta x \left({\cos{\pi k\over n}\over n\sin{\pi k\over n}}\right)^2+O(x^2)\right]\cr}}
This is manifestly of $O(c^0)$, for any $c$. Notably, it is spin-independent. If the CFT contains no Virasoro primary operators with conformal dimension less than or equal to one, \sumry\ is the full contribution of $\Oc$ to this order in $x$.

Happily, \sumry\ also admits a tidy analytic continuation $n\rar 1$, using the method of \CalabreseHE. This yields the {\it entanglement} entropy contribution of this spinning primary, $S_{EE}^{(\Delta,s)}=\lim_{n\rar 1}S_n^{(\Delta,s)}$. We find
\eqn\sumr{S_{EE}^{(\Delta,s)}\Big|_{x\ll1} = \left({x\over 4}\right)^{2\Delta}{\sqrt{\pi}\over 4}\left[{\Gamma(2\Delta+1)\over\Gamma(2\Delta+{3\over 2})} + x\Delta {\Gamma(2\Delta+2)\over\Gamma(2\Delta+{5\over 2})} +O(x^2)\right]}
The leading term in $x$ was known for spinless operators only \refs{\HeadrickZT,\CalabreseHE}. The next-to-leading term is new. 

\newsec{R\'enyi entropy in higher spin theories II: Short interval expansion}
We return to the calculation of the ground state two interval R\'enyi entropy of a higher spin CFT. To leading and next-to-leading order in small $x$, the contributions to the R\'enyi entropy from a given pair of higher spin currents $(J^{(s)},\overline{J}^{(s)})$ can be simply read off from \sumry.\foot{For those who wish to see more details on the spectrum of quasiprimaries of $\C^n$ when $\C$ possesses higher spin symmetry, in Appendix B we give an explicit accounting of these fields by studying the decomposition of the relevant character into sl(2,$\IR$) highest weight representations.} The holomorphic current has $(h,\hb)=(s,0)$, and the antiholomorphic current with $(h,\hb)=(0,s)$ contributes identically. Denoting their combined contribution as $S_n^{(s)}$ so as to make contact with the bulk calculation involving a spin-$s$ gauge field, we find
\eqn\olf{\eqalign{&S^{(s)}_n\big|_{x\ll 1}  \approx  {2x^{2s}\over 1-n}\sum_{k=1}^{n-1}(n-k) \left({1\over 4n^2\sin^{2}{\pi k\over n}}\right)^{2s}\left[1+sx+sx\left({\cos{\pi k\over n}\over n\sin{\pi k\over n}}\right)^2+O(x^{2})\right]  }}
Our charge is to show that 
\eqn\rba{S^{(s)}_n\big|_{x\ll 1}=S^{(s)}_n(\cM)\big|_{\rm 1-loop} }
where $S^{(s)}_n(\cM)\big|_{\rm 1-loop}$ is the 1-loop determinant on handlebody geometries $\cM$ for a spin-$s$ gauge field, evaluated for $x\ll1$.

\subsec{Higher spin fluctuations on handlebody geometries}
In Section 2 we recalled the formula for the 1-loop partition function on a handlebody geometry $\cM=\IH^3/\G$, which we reproduce here for convenience:
\eqn\ola{Z_{\IH^3/\G} = \prod_{\g\in {\cal P}} \left(Z_{\IH^3/\IZ}(q_{\g})\right)^{1/2}}
The legend is as follows. The elements\foot{A word on notation: these $\g$ are not accessory parameters!} $\g$ belong to $\cP$, a set of representatives of the primitive conjugacy classes of $\G$. ``Primitive'' means that $\G$ cannot be written as $\b^n$ for $\b\in \G$ and $n>1$. The eigenvalues of the element $\g$ are parameterized as $q_{\g}^{\pm 1/2}$ with $|q_{\g}|<1$, and are functions of the geometric data of $\cM$.

Now we note that the 1-loop partition function for spin-$s$ gauge fields on the solid torus $\IH^3/\IZ$ is known \DavidXG; plugging this into \ola, we find
\eqn\olb{Z_{\IH^3/\Gamma}^{(s)} = \prod_{\g\in {\cal P}} \prod_{n=s}^{\infty}{1\over |1-q_{\g}^n|}}
The loop counting parameter is $1/k\sim 1/c$. Our conjecture thus implies that \olb\ yields the $O(c^0)$ contribution to the R\'enyi entropy from a pair of spin-$s$ chiral currents in a dual CFT:
\eqn\olc{S_n^{(s)}(\cM)\big|_{\rm 1-loop} = -{1\over 1-n}\sum_{\gamma \in {\cal P}}\sum_{n=s}^{\infty}{\log |1-q_{\g}^n|} }
It also implies that the set of elements $\g\in\G$ over which we sum in \olc\ is the same as in pure gravity, hence so are the eigenvalues $q_{\g}$. The algorithm for computing the $q_{\g}$ in a short interval expansion, and performing the sum over $\g\in {\cal P}$, was presented in \BarrellaWJA. At 1-loop, only the cross ratio appears in the R\'enyi entropy, so ``short interval" means small $x$. 

Let us recall the salient features of the computations of \BarrellaWJA. For $n>2$, the sum over $\g\in {\cal P}$ in \olc\ is infinite. However, in a small $x$ expansion, only a finite subset of the representatives (``words'') $\g$ of the primitive conjugacy classes of $\G$ contribute at a given order in $x$. \BarrellaWJA\ systematically organized how the eigenvalues $q_{\g}$ scale with $x$, depending on the nature of the word $\g$.

All words are constructed from the Schottky generators $L_i$, where $i=1,\ldots,g=n-1$. A basis of words that can be used to construct all other words is spanned by the so-called ``consecutively decreasing words'' (CDWs), which take the form
\eqn\ole{\g_{k,m} = L_{k+m}L_{k+m-1}\ldots L_{m+1}}
They are built from $k$ generators $L_i$ ordered in obvious reference to the nomenclature. At length $k$, there are $n-k$ unique CDWs indexed by $m$ in \ole, where $0\leq m \leq n-1-k$. At $k=1$ the set of $n-1$ CDWs is just given by the $L_i$ themselves; at $k=2$, we have $\lbrace L_2L_1, L_3L_2,\ldots,L_{n-1}L_{n-2}\rbrace$; and so on.  

Having laid this groundwork, we list the facts we will need:
\vs
1. The eigenvalues of $\g_{k,m}$ are $m$-independent.
\vs
2. All non-CDWs can be formed by gluing CDWs together; following \BarrellaWJA, we call these more general words $p$-CDWs. 
\vs
3. In a small $x$ expansion, the larger of the two eigenvalues of $p$-CDWs,  $q_{\g}^{-1/2}$, scales as $x^{-p}$ to leading order in $x$, and is real. Thus, small $x$ means small $q_{\g}$.\vs

At small $q_{\g}$, 
\eqn\old{\eqalign{S_n^{(s)}(\cM)\big|_{\rm 1-loop}& =  {1\over 1-n}\sum_{\gamma \in {\cal P}}\Re[q_{\g}^s+O(q_{\g}^{s+1})]}}
Therefore, the contribution of a $p$-CDW to the 1-loop R\'enyi entropy from a spin-$s$ current starts at $O(x^{2sp})$.  For the sake of comparison with CFT, we content ourselves with a presentation of the leading and next-to-leading order contributions in $x$. As the higher spin fields have $s\geq 2$, we can focus exclusively on the CDW contribution. 

Employing the algorithm of \BarrellaWJA\ to compute $q_{\g}$ at small $x$, we find the larger of the two CDW eigenvalues to be
\eqn\ole{q_{\g}^{-1/2} = -{4n^2\sin^2{\pi k\over n}\over x}+2\cos^2{\pi k\over n}+2n^2\sin^2{\pi k\over n}+O(x)}
Summing over the $n-k$ inequivalent CDWs of length $k=1,\ldots,n-1$ and their inverses in \old, we obtain the desired result,
\eqn\olfa{\eqalign{&S_n^{(s)}(\cM)\big|_{\rm 1-loop} \approx  {2x^{2s}\over 1-n}\sum_{k=1}^{n-1}(n-k) \left({1\over 4n^2\sin^{2}{\pi k\over n}}\right)^{2s}\left[1+sx+sx\left({\cos{\pi k\over n}\over n\sin{\pi k\over n}}\right)^2+O(x^{2})\right]  }}
This is precisely \olf: in other words,
\eqn\qsg{S^{(s)}_n\big|_{x\ll 1}=S^{(s)}_n(\cM)\big|_{\rm 1-loop} }
Subsequent holographic calculations \BeccariaLQA\ of higher spin gauge field contributions to entanglement entropy have reproduced the $n\rar 1$ limit of \olfa, which is twice equation \sumr\ with $\Delta=s$.

\subsec{Comments}
This match gives an appealing flavor to the bulk result: the individual terms in the sum over CFT operators are in one-to-one correspondence with CDW contributions to the expansion of the bulk determinant. It is highly sensitive to the existence and use of the correct saddle point geometry.

Beyond lowest orders, however, it becomes increasingly cumbersome to understand the CFT origin of terms in the bulk determinant, where there is mixing among conformal families. It would be nice to understand, as we do for the case of 1-loop gravity on the solid torus, how to re-organize the CFT conformal block decomposition such that the match to gravity is manifest, or vice-versa. 

\newsec{Chiral asymmetry and $c_L\neq c_R$ CFTs }

For the remainder of the paper, we will stay at leading order in large central charge, so what follows applies to CFTs with or without an extended chiral algebra. We now allow unequal left and right central charges, $c_L\neq c_R$. Our goal is to answer the question, ``are the R\'enyi entropies similarly universal at large total central charge?'' The answer is yes, and the formula is as follows: at leading order in large $(c_L,c_R)$,
\eqn\rsa{{\p S_n\over \p z_i} = {n\over 6(n-1)}(c_L\g_{L,i}+c_R\g_{R,i})}
where we have introduced left and right accessory parameters $(\g_L,\g_R)$, each of which is related to the respective chiral, semiclassical Virasoro conformal block as in \inl. For only, say, $c_L$ large, the formula is again \rsa\ at leading order, but with only the left piece. 

To explain the uncomplicated origin of this formula, we first need to generalize our results to chirally asymmetric states, in which left- and right-movers behave differently. We begin by taking $c_L=c_R=c$ and generalize in due course.

\subsec{Chirally asymmetric states}
It is well-known that in general, the accessory parameters appearing in the Schottky uniformization are complex. We can parameterize them in terms of independent left and right accessory parameters, $\g_L$ and $\g_R$. Like 2d complex coordinates, we regard these as generically independent variables. In the background of the heavy operators, the holomorphic and anti-holomorphic stress tensors need not be equal as a function of their respective coordinates, so the uniformization equations
\eqn\csa{\eqalign{\psi''(z)+T(z)\psi(z)&=0\cr
\psib''(\zb)+\overline{T}(\zb)\psib(\zb)&=0\cr}}
are independent. (We make the associations ``left'' $\leftrightarrow$ ``unbarred'', and ``right'' $\leftrightarrow$ ``barred''.) Indeed, the solutions to these two equations parameterize the most general complex solution to the Liouville equation \HarlowNY.

Based on this factorization, the formula for the R\'enyi entropy in such a state is 
\eqn\cha{{\p S_n\over \p z_i} = {cn\over 6(n-1)}(\g_{L,i}+\g_{R,i}) = {cn\over 3(n-1)}\Re[\g_i]}
This is manifestly real, as required. We will give an explanation of \cha\ from first principles in the next subsection. 

Let us first discuss an instructive example, namely, the case of a CFT on a circle of unit radius at nonzero temperature, $T=\b^{-1}$, and chemical potential for angular momentum, $\Omega$. The presence of nonzero $\Omega$ leads to unequal left and right temperatures, $(\b_L^{-1},\b_R^{-1})$. In Euclidean signature, the CFT lives on a torus with complex $\t$,
\eqn\chac{\t = {i\b(1+i\Omega_E)\over 2\pi}}
where we have defined $\Omega_E$ to be real via the analytic continuation $\Omega=i\Omega_E$. In terms of $\b_L$ and $\b_R$, 
\eqn\clb{i\b_L = 2\pi\t~, ~~ i\b_R = -2\pi\taub}
The coordinates thus obey the identification $(z,\zb) \sim (z+2\pi, \zb+2\pi)\sim (z+i\b_L, \zb+i\b_R)$. 

We now want to compute the single-interval R\'enyi entropy on this torus. This was done in \BarrellaWJA\ for the case of purely imaginary $\t$. Our first step is to uniformize this branched cover of a torus with complex $\t$, after which we apply \cha. The uniformization equations \csa\ are essentially two copies of the equations used to uniformize the branched cover of the rectangular torus, one each on the left and right. Accordingly, one can simply read off the desired results from \BarrellaWJA, and in our conventions, these equations read 
\eqn\cld{\eqalign{\psi''(z) + \sum_{i=1}^2\left({6h_{\Phi}\over c}~\wp_L(z-z_i)-\g_L(-1)^{i+1}\zeta_L(z-z_i)+\delta_L\right)\psi(z)=0\cr
\psib''(\zb) + \sum_{i=1}^2\left({6\hb_{\Phi}\over c}~\wp_R(\zb-\zb_i)-\g_R(-1)^{i+1}\zeta_R(\zb-\zb_i)+\delta_R\right)\psib(\zb)=0\cr}}
Here, $\wp_{L}$ and $\zeta_{L}$ are shorthand for the Weierstrass elliptic and zeta functions, respectively, with periodicities fixed with respect to the $z$ coordinate, and similarly for $\wp_R$ and $\zeta_R$ with periodicities fixed with respect to $\zb$. (e.g. $\wp_L(z-z_i)$ is periodic under $z \sim z+2\pi \sim z+i\b_L$.) For a trivial monodromy around the spatial cycle, $\delta_{L}=\delta_R=1/8$; for trivial monodromy around the thermal cycle, $\delta_L= -(\pi T_L)^2/ 2$ and $\delta_R=-(\pi T_R)^2/2$. 

We can solve these equations analytically in the entanglement limit $n\rar 1+\eps$. In the case where we trivialize the monodromy around the thermal circle, we find
\eqn\hna{\eqalign{\lim_{n\rar 1}\g_L &= -\pi T_L\coth(\pi T_Lz_{21})~\eps+O(\eps^2)\cr
\lim_{n\rar 1}\g_R &= -\pi T_R\coth(\pi T_Rz_{21})~\eps+O(\eps^2)\cr}}
where $z_{21}=z_2-z_1$. To obtain the entanglement entropy, we follow \BarrellaWJA\ in assuming that \cha\ holds even on the torus.\foot{As noted in the introduction, it would be very helpful to {\it prove} that the procedure utilized here and in \BarrellaWJA\ for computing torus R\'enyi entropy is valid. One may be able to prove \cha\ from CFT by studying torus conformal blocks. \PiatekIFA\ and references therein may be useful. Relatedly, we expect that, in analogy to the case of two intervals on the plane \zamo, the torus uniformization equations \cld\ can be understood as null decoupling equations obeyed by Liouville three-point functions involving two heavy operators and one light, degenerate operator with a null state at level two. In any case, the arguments of \refs{\HartmanMIA, \FaulknerYIA} are sufficient to ensure the validity of our results (6.8) and (6.13) below in the decompactification limit in which the CFT lives on the cylinder.} Integrating using \cha, we find the entanglement entropy at leading order in $c$,
\eqn\clf{S_{EE} = {c\over 6}\log\left[\sinh(\pi T_L z_{21})\sinh(\pi T_R z_{21})\right]}
where we have suppressed an integration constant. We find an unsurprising universality: at large central charge in a spatially compact CFT with $T_L\neq T_R$, the finite size of the circle is invisible in the leading approximation to the single interval entanglement entropy. The same universality was found in \BarrellaWJA\ when $T_L=T_R$. \clf\ agrees with the universal part of the bulk result obtained using the covariant generalization of the Ryu-Takayanagi formula in the rotating BTZ black hole background \HubenyXT.

When the interval spans the entire circle, $z_{21}=2\pi$, and taking the high temperature limit $(T_L,T_R)\rar\infty$, one recovers the Cardy formula for rotating black hole entropy as applied to theories with $c_L=c_R$,
\eqn\clg{S_{\rm BTZ} = {c\over 3}{\pi^2}(T_L+T_R)}
Of course, the Cardy formula equally applies to CFTs with $c_L\neq c_R$, and our example suggests that the formula \cha\ for R\'enyi entropies enjoys an equally straightforward generalization. Let us turn to this case now.

\subsec{Chirally asymmetric theories}
We now allow $c_L\neq c_R$. To begin, we consider theories in which $(c_L,c_R)$ are both large. The correct formula for the R\'enyi entropy, to leading order in the central charges, is
\eqn\chcc{{\p S_n\over \p z_i} = {n\over 6(n-1)}(c_L\g_{L,i}+c_R\g_{R,i})}
Alternatively, writing $\g=\Re[\g]+i\Im[\g]$,
\eqn\chcd{{\p S_n\over \p z_i} = {n\over 6(n-1)}\big((c_L+c_R)\Re[\g]+i(c_L-c_R)\Im[\g]\big)}
These formulae reduce to \cha\ when $c_L=c_R$. We see that only in a CFT with $c_L\neq c_R$ does the chiral asymmetry of a state affect the R\'enyi entropy to leading order; conversely, only in chirally asymmetric states does the R\'enyi entropy receive an anomalous contribution.

It takes little work to derive \chcc. The key point is that the conformal block decomposition of the twist field correlators holomorphically factorizes for each conformal family. The left and right Virasoro conformal blocks are only functions of their respective central charges, and they are the generating functions for their respective accessory parameters. When $(c_L,c_R)$ are both large, the arguments of \HartmanMIA\ apply independently on the left and right. The result is \chcc. We have derived this formula in the ground state or states related by conformal transformations which need not have $\g_L = \g_R$ (e.g. at finite temperatures $T_L\neq T_R$), though we expect it to hold in more general states, as stressed earlier.\foot{\chcc\ and \chcd\ make the R\'enyi entropy appear complex. This should be viewed on the same footing as the rotating BTZ black hole entropy as computed in Euclidean signature: in that case, the analytic continuation of the chemical potential $\Omega$ ensures that the Virasoro zero modes in the original, Lorentzian theory are real. We thank Matt Headrick for pointing out this analogy.} 

Perhaps this is surprising: CFTs with $c_L\neq c_R$ suffer from a gravitational anomaly, which implies non-conservation of the stress tensor controlled by $c_L-c_R$. This can be traded for an anomaly under local Lorentz transformations by the addition of a local counterterm to the CFT action. Either way, one might wonder whether the formulae for R\'enyi entropy are affected by the anomaly. Regardless of the relative values of $(c_L,c_R)$, we can continue to define R\'enyi entropies via path integrals of singular branched covering spaces. These can always be traded for correlation functions of twist fields in the orbifold CFT, whereupon the previous paragraph applies.\foot{We should note that we are not the first to consider entanglement entropy in theories with $c_L\neq c_R$: for a single interval in the ground state, the result following from \chcc\ was derived in the original work of \HolzheyWE. We also note \SunUF, which derived some holographic results for single interval entanglement. While that author's results for the ground state and static BTZ are consistent with \chcc, he appears to get a different result for rotating BTZ. More investigation in the bulk is warranted.}

We can also relax our assumption slightly, and consider the regime in which only one central charge is large: $c_L+c_R\gg 1$, but $c_L\gg c_R\sim O(1)$. In this case, the R\'enyi entropy still exhibits universal behavior: to leading order in $c_L$, we have a chiral half of \chcc, 
\eqn\chcca{{\p S_n\over \p z_i} = {nc_L\over 6(n-1)}\g_{L,i}}
The argument is again brief: the left conformal block exponentiates and is dominated by the vacuum block, but the right conformal block is generic and does not scale with any power of $c_L$. So the entire leading contribution comes from the left; at subleading orders, the right conformal block mixes with the $O(c_L^0)$ corrections to the result \chcca. Note that this logic applies to chiral CFTs of the sort considered in the context of topologically massive gravity, to which we will turn shortly \refs{\LiDQ,\MaloneyCK}.

\chcc\ looks like many other formulae in 2d CFT, namely the Cardy formula. Indeed, we must be able to recover the Cardy formula from the single interval entanglement entropy on the torus. Combining \chcc\ with the results of the previous subsection yields this entanglement entropy,
\eqn\clm{S_{EE} = {c_L\over 6}\log\sinh(\pi T_Lz_{21}) + {c_R\over 6}\log\sinh(\pi T_Rz_{21}) }
Again, we see no finite size effects. In the high temperature limit when $z_{21}=2\pi$, we recover the Cardy formula,
\eqn\cln{S_{\rm BTZ} = {\pi^2\over 3}(c_LT_L+c_RT_R)}
So \chcc\ passes a necessary check: it gives the correct density of states of a general CFT on a torus.

\subsec{Spinning twist fields from gravitational anomalies}
While the anomaly does not affect the prescription, it does have (at least) one interesting effect. In particular, it forces the twist fields to acquire a nonzero spin, proportional to $c_L-c_R$. The formula \chcc\ tells us that this spin only affects the R\'enyi entropy in chirally asymmetric states.

To see this, it suffices only to note that the determination of the twist field scaling dimensions $(h_{\Phi},\hb_{\Phi})$ is a chiral calculation. The dimensions of these Virasoro primary operators are fixed by conformal Ward identities of the holomorphic and anti-holomorphic stress tensors. Retracing the original arguments of Calabrese and Cardy, it is easy to derive

\eqn\spe{\eqalign{h_{\Phi} = {c_L\over 24}\left(n-{1\over n}\right)~, ~~\hb_{\Phi}= {c_R\over 24}\left(n-{1\over n}\right) }}
The twist field has spin controlled by the anomaly coefficient $c_L-c_R$. 

Even though $h_{\Phi}\neq \hb_{\Phi}$, this only affects the R\'enyi entropy in chirally asymmetric states because the chiral semiclassical Virasoro blocks depend only on the combination $h_{\Phi}/c_L$ or $\hb_{\Phi}/c_R$. We see in \clm, for example, that in the absence of rotation, $T_L=T_R$ and only the total central charge appears. 

Interestingly, the anomaly also manifests itself when the intervals are not purely spatial.  Consider the case of one interval in the ground state with endpoints $(z_1,z_2)=(0,v)$, with $v$ complex. When $c_L=c_R=c$, we can rotate our axes to put the interval on the real line. However, in a theory with an anomaly under Lorentz transformations, observables are not invariant under this rotation. Using conformal invariance to fix the twist field two-point function, the R\'enyi entropy is
\eqn\spg{\eqalign{S_n &= \left(1+{1\over n}\right)\Big[{c_L\over 12}\log v+{c_R\over 12}\log\bar v\Big]\cr}}
If we write $v=\ell e^{i\theta}$, we have
\eqn\clq{S_n = \left(1+{1\over n}\right)\left[{c_L+c_R\over 12}\log {\ell\over \eps} + {c_L-c_R\over 12}i\theta\right]}
This phase signals the presence of the anomaly.

\subsec{A holographic perspective}
The stress tensor sector of a holographic CFT with $c_L\neq c_R$ is dual to topologically massive gravity (TMG). This theory augments pure 3d gravity with a propagating spin-2 degree of freedom. In metric language, the Lorentzian theory has an action
\eqn\tma{S_{\rm TMG} = {1\over 16\pi G}\int d^3x \sqrt{-g} (R-2\Lambda) + {\alpha} S_{CS}[\Gamma] + S_{\rm bndy}}
where 
\eqn\tmb{S_{CS}[\Gamma] =\int\Tr\left(\G \wedge d\G + {2\over 3}\G\wedge \G \wedge \G\right)}
$S_{\rm bndy}$ denotes a boundary action. The second term is the gravitational Chern-Simons term for the Christoffel connection, written here as a connection one-form with two implicit matrix indices, $\G = \G_{\rho}dx^{\rho}$. With standard asymptotically AdS boundary conditions, the Virasoro central charges are
\eqn\tmbb{c_L+c_R={3\ell\over G}~, ~~ c_L-c_R = 96\pi\alpha}
This theory is classical as long as the total central charge is large. Bulk diffeomorphisms leave the action invariant up to a boundary term that reflects the CFT gravitational anomaly \KrausZM. 

At least classically, TMG can be written in a first-order formulation. The action is given by a sum of sl(2,$\IR$) Chern-Simons actions with unequal levels and a Lagrange multiplier term enforcing the zero torsion constraint:
\eqn\tmd{S_{\rm TMG} = {k_L\over 4\pi}S_{CS}[A] - {k_R\over 4\pi}S_{CS}[\Ab] +  {k_L-k_R\over 8\pi}\int \Tr\left(\chi\wedge (F-\Fb)\right)}
$\chi$ is a Lagrange multiplier that enforces zero torsion, $F=\overline{F}$ \CarlipQH. The levels $k_{L,R}$ relate to the CFT central charges as
\eqn\tmda{c_L=6k_L~, ~~ c_R=6k_R}
In this language, $\chi$ is the new degree of freedom. Without it, the theory is a pure topological Chern-Simons theory, which is not TMG.

From the point of view of the AdS/CFT correspondence, we identify CFT R\'enyi entropy $S_n$ with the Euclidean bulk path integral over manifolds with the desired replica surface at conformal infinity. Our CFT result \chcc\ implies that the on-shell TMG action $S_{TMG}$, evaluated on R\'enyi handlebody solutions $\cM$ characterized by accessory parameters $(\g_{L,i},\g_{R,i})$ coming from the Schottky uniformization of $\p\cM=\Sigma$, is given by the factorized expression
\eqn\cpb{{\p S_{\rm TMG}[\cM]\over \p z_i} = {n\over 6}(c_L\g_{L,i}+c_R\g_{R,i})}
This result assumed that both $(c_L,c_R)$ are large; if only, say, $c_L$ is large, we should only take the $c_L$ term seriously at leading order. This scenario occurs at the much-studied chiral point, $c_R=0$ \refs{\LiDQ,\MaloneyCK}. 

It would be nice to rigorously prove \cpb. Here, we put forth some bulk arguments as to how one might do so. 

We will employ the first-order formalism \tmd. This theory admits all the solutions of pure AdS gravity ($F=\Fb=\chi=0$), as well as new solutions that are not flat connections. Evaluated on-shell on any solution of the theory, the Lagrange multiplier term vanishes. So the full on-shell action is given by the on-shell Chern-Simons actions, appropriately renormalized.\foot{We are evaluating the classical action on an Einstein saddle point, so we can ignore the subtleties regarding the relation between Chern-Simons theory and quantum gravity. We also note that this argument has been used in the literature before. For instance, evaluated on real Euclidean metrics, the action of Euclidean TMG reduces to that of pure Chern-Simons theory \MaloneyCK.}

We want to evaluate \tmd\ on the handlebody solutions, $\cM$. By ``solution'' we mean the pair of flat connections $(\cA,\cAb)$ which maps back to a given handlebody in metric language.\foot{The evaluation of the on-shell Einstein action for the handlebodies was done in metric language in \FaulknerYIA. As a warm-up to the TMG case discussed here, it would be instructive to translate that procedure to Chern-Simons language \JJ.} 
 As in the metric formulation, we must add boundary terms linear in $k_L$ or $k_R$ to render the result finite. A natural outcome is that the left and right Chern-Simons actions are renormalized individually. Denoting the renormalized actions with a hat, we can then write the on-shell action as
\eqn\cpa{\widehat{S}_{\rm TMG}[\cM] = {k_L\over 4\pi}\widehat{S}_{CS}[\cA]-{k_R\over 4\pi}\widehat{S}_{CS}[\cAb]}
and \cpb\ then implies that 
\eqn\cpc{{\p \widehat{S}_{CS}[\cA]\over \p z_i} = {4\pi n\g_{L,i}}~, ~~ {\p \widehat{S}_{CS}[\cAb]\over \p \zb_i} = -{4\pi n\g_{R,i}}}

This kind of holomorphic factorization of classical actions has some precedent in 3d gravity. From the point of view of classical gravity as a pair of decoupled Chern-Simons theories, this factorization is a sensible expectation. \cpc\ says that the holomorphic Chern-Simons action is determined by the residue of the holomorphic stress tensor $T(z)$ in the background of the heavy twist fields as we take $z\rar z_i$, and an identical statement holds on the anti-holomorphic side. The boundary monodromy problem maps to a bulk holonomy problem, and the two bulk sectors are completely decoupled in this respect, just as they are in the determination of asymptotic symmetries. 

When the bulk has the topology of a solid torus with modular parameter $\t$ -- corresponding to the BTZ black hole or $SL(2,\IZ)$ images thereof -- factorization of the classical contribution of a fixed saddle is known to hold \MaldacenaBW. We are holographically computing a CFT partition function in the canonical ensemble, which takes the form $Z = \Tr(q^{L_0-{c_L\over 24}}\overline{q}^{\Lb_0-{c_R\over 24}})$ where $q=e^{2\pi i \t}$. The semiclassical bulk action yields the leading term in this trace. Higher genus examples of factorized tree-level partition functions in Einstein gravity are given in \YinGV.  

Finally, recalling that our CFT result applies even in the presence of a higher spin symmetry, we should address proposals for theories of topologically massive higher spins \refs{\ChenVP, \BagchiVR}. Their actions are quite like \tmd: a pair of Chern-Simons actions with $k_L\neq k_R$ for gauge fields valued in some higher spin algebra $\cG$, accompanied by one Lagrange multiplier for each member of the Cartan subalgebra of $\cG$. The latter enforce the vanishing of all components of the $\cG$-valued torsion. Our problem is again reduced to evaluating Chern-Simons actions on the sl(2,$\IR$)~$\times$ sl(2,$\IR$) handlebody solutions, which yields the same result as in ordinary TMG.

\newsec{Discussion}
To conclude, we have shown that the formulae of \refs{\HartmanMIA,\FaulknerYIA} for multiple interval R\'enyi entropies in 2d CFTs at large central charge generalize straightforwardly to a very wide class of theories. This class includes nearly all CFTs which have been considered in the context of a holographic correspondence. In the ground state, this amounts to an extension of the Ryu-Takayanagi formula for holographic entanglement entropy to theories beyond higher-derivative-corrected Einstein gravity.

In the main text, we alluded to various next steps that we would like to see carried out. Let us close with some related ideas. \vs

\noindent \bul {\it $W_N$ minimal models\vs}
The $W_N$ minimal model holography proposals \refs{\GaberdielPZ,\GaberdielKU} do not fit neatly into our discussion. While they possess the right algebras, they either fail to be unitary \refs{\GaberdielKU, \PerlmutterDS} or spectrally sparse \refs{\GaberdielPZ,\PapadodimasPF, \GaberdielCCA} depending on the precise large $c$ limit taken. In the better-studied 't Hooft limit, these theories have a near-continuum of light states below the gap that grows exponentially with $c$. These states pollute the large $c$ factorization properties of the theory and wash out any bulk phase transitions \refs{\GaberdielCCA,\BanerjeeAJ}.  It is fair to say that the precise holographic dictionary with the Vasiliev theory has yet to be fully understood. In any case, we cannot yet say anything about the ground state R\'enyi entropies of these theories. \vs
\noindent \bul {\it Backgrounds with higher spin charge\vs} 

An obvious limitation of this work is that we have nothing to say about R\'enyi entropies in states of higher spin CFTs with nonzero higher spin charge, chemical potential or both. We expect that higher spin symmetry may be powerful enough to determine the answer completely. There has been progress in evaluating bulk partition functions in higher spin theories for solutions with solid torus topology and nonzero higher spin charge \refs{\KrausDS, \BanadosUE, \deBoerGZ}, and we may hope to apply those lessons to the bulk R\'enyi calculations. Perhaps this should involve classifying quotients of AdS$_3$ by generic elements of the bulk higher spin gauge algebra. 

The bulk solution of this problem is bound to be interesting. There are already two proposals  \refs{\AmmonHBA,\deBoerVCA}\ for computing single interval entanglement entropy in generic backgrounds of sl(N,$\IR)~\times~$sl(N,$\IR$) Chern-Simons theories.\foot{They appear to be distinct, but give the same answers for the handful of backgrounds in which they have been evaluated explicitly. This coincidence has yet to be understood.} These prescriptions are quite natural from a Chern-Simons perspective: they are given in terms of Wilson lines for the bulk connection. An extension to R\'enyi entropies and to multiple intervals begs to be discovered. It would be satisfying to forge a link between these higher spin R\'enyi entropies and the full family of topological observables in Chern-Simons theories.\foot{We thank Alejandra Castro, Nabil Iqbal, Juan Jottar and Wei Li for discussions on these topics.}\vs

\noindent \bul {\it R\'enyi entropy in moduli space\vs}
It is interesting to consider theories with a moduli space at large central charge. In some regions of the moduli space, the conclusions herein may hold, but not in other regions where the CFT fails to obey our assumptions. It would be instructive to try to microscopically and quantitatively understand how the R\'enyi entropy changes as one moves in such a moduli space. 

To this end, a family of CFTs that one might consider is the family of large $\N=4$ superconformal field theories, recently considered in a higher spin light \GaberdielVVA. A specific class of coset CFTs has been identified that admits a large $c$ limit in which the theory is conjecturally dual to a supersymmetrized, non-Abelian Vasiliev theory of 3d higher spins. On the other hand, type IIB supergravity on AdS$_3\times S^3\times S^3 \times S^1$ has the same large $\N=4$ supersymmetry, though no widely accepted proposal exists for its CFT dual \GukovYM. Nevertheless, the arguments of \GaberdielVVA\ imply that the high energy limit of type IIB string theory on the above background contains a Vasiliev sector. There should be a family of dual CFTs that mirrors this interpolation; while it has yet to be found, there is promising quantitative evidence that it should exist. 

At the higher spin point where there {\it is} a specific conjectured coset CFT dual to the Vasiliev theory, the CFT enjoys a $W$ symmetry, labeled $s\tilde{\Wc}_{\infty}^{(4)}[\g]$, that includes eight fields at each spin $s=3/2,2,5/2,\ldots$, in addition to seven spin-1 currents. To leading order, this spectrum in itself would lead to the same large $c$ R\'enyi entropies as $W_N$ and $W_{\infty}[\l]$; however, like the bosonic $W_N$ minimal models, the coset also contains a tower of matter primaries that includes (too) many light primaries. These should lift upon passage to the string theory regime. We hope that this case might give a fully tractable instance of holography along the full moduli space, in which we can study the breakdown of classical entanglement entropy explicitly. \vs

\noindent \bul {\it R\'enyi entropy in TMG\vs}

The extension of the Ryu-Takayanagi prescription to general TMG backgrounds is an open problem. Whatever the correct prescription, it should reproduce the result \clm\ when applied to the rotating BTZ black hole background. One should be able to apply recent techniques involving regulated cones to this problem. These issues are currently being taken up in \tmgp. 

One of the mysteries of TMG is why the rotating BTZ black hole entropy picks up a contribution proportional to the inner horizon area \refs{\SaidaEC,\KrausVZ,\SahooVZ,\TachikawaSZ,\SolodukhinAH}. That this must be true follows from Cardy's formula, but we lack a geometric picture for this. Hopefully, entanglement calculations can shed some light; indeed, it seems that the nonzero spin \spe\ of the twist field in the dual CFT makes an essential difference \tmgp. 

We also note that the on-shell action methods of \refs{\BanadosUE, \deBoerGZ} fix $k_L=k_R$, but appear to admit a generalization to $k_L\neq k_R$. It would also be worthwhile to investigate whether their formulae admit straightforward extensions to higher genus, which would allow a direct calculation of \cpb.

Besides AdS and its quotients, the most intriguing backgrounds of TMG are warped AdS and its quotients \AnninosFX. The results in Section 6 do not apply to warped AdS, nor to warped CFTs \DetournayPC: we do not yet understand warped CFT conformal blocks, nor is warped AdS a quotient of ordinary AdS. Some work on entanglement entropy in warped AdS has been done on the gravity side \AnninosNJA, but not in a TMG context. It would be quite interesting to investigate entanglement entropy in warped CFTs. 

\vskip .3in

\noindent
{ \bf Acknowledgments}

\vskip .3cm
We thank Martin Ammon, Taylor Barrella, Andrea Campoleoni, John Cardy, Alejandra Castro, Stephane Detournay, Matthias Gaberdiel, Matt Headrick, Nabil Iqbal, Juan Jottar, Per Kraus, Albion Lawrence, Wei Li, Mukund Rangamani, Evgeny Skvortsov and Erik Tonni for very helpful discussions; Alejandra Castro, Nabil Iqbal and an anonymous referee for comments on the manuscript; and especially Tom Hartman for crucial feedback and conversations.  We also wish to thank the Isaac Newton Institute (Cambridge), CERN, and the Max Planck Institute for Gravitational Physics (Potsdam-Golm) for hospitality while this work was in progress. The author has received funding from the European Research Council under the European Union's Seventh Framework Programme (FP7/2007-2013), ERC Grant agreement STG 279943, “Strongly Coupled Systems”.   

\appendix{A}{The $W_3$ algebra}
The quantum $W_3$ algebra has commutation relations
\eqn\wta{\eqalign{
[L_{n},L_{m}]&=(n-m)L_{n+m}+{c\over 12}n(n^2-1)\delta_{m,-n}\cr
[L_{n},W_{m}]&=({2n}-m)W_{n+m}\cr
[W_n,W_m]&={c\over 360}n(n^2-1)(n^2-4)\delta_{m,-n}+{16\over 22+5c}(n-m)\Lambda_{m+n}\cr
&+(n-m)\left({1\over 15}(m+n+3)(m+n+2)-{1\over 6}(m+2)(n+2)\right)L_{m+n}\cr}}
where
\eqn\wtb{\Lambda_{m}=\sum_{p\in\IZ}:L_{m-p}L_{p}:-{3\over 10}(m+3)(m+2)L_{m}}
The notation $: \,:$ denotes normal ordering, and all generators are integrally moded. The first line is the Virasoro algebra; the second line is the statement that the $W$ current is Virasoro primary; and the third line is the juicy part. Note the non-linearities, down by $1/c$, permitting a definition of the wedge algebra spanned by modes $(L_{m},W_n)$ with $|m|\leq1, |n|\leq 2$.

\appendix{B}{Quasiprimaries in tensor product higher spin CFTs}
One can easily understand the spectrum of quasiprimaries of a product theory $\C^n$ by branching the characters of $\C^n$ into characters of sl(2,$\IR$) highest weight representations. The latter is given by
\eqn\chl{\chi_h^{sl(2,\IR)} = {q^h\over 1-q}}
for a quasiprimary of highest weight $h$. For present purposes, we are interested in isolating quasiprimaries made solely of current modes. A current of spin $s$ is a level $s$ descendant of the vacuum, so we want to branch $n$ powers of the vacuum character. 

To be completely explicit, we focus on the case where $\C$ has $W_3$ symmetry. Decomposition of the $n$'th power of the $W_3$ vacuum character, $\chi_3$, will count all quasiprimaries $\Och$ made of only current modes. To isolate those built from at least one $W$ current, we should subtract the pure Virasoro part, $(\chi_{Vir})^n$. Writing
\eqn\chm{(\chi_3)^n- (\chi_{Vir})^n= {1\over 1-q}\sum_hd_W(h)q^h}
we obtain a generating functional for quasiprimaries of $\C^n$ involving at least one $W$ current:
\eqn\chn{\eqalign{\sum_hd_W(h)q^h &= (\left(\chi_3)^n- (\chi_{Vir})^n\right)(1-q)\cr
&\approx n q^3 + n^2 q^5 + {n(3n+1)\over 2}q^6+O(q^7)}}
At level three, these are the $W$ currents living on any of $n$ sheets. At level five, we find quasiprimaries with $T$ and $W$ living on separate sheets; there are $n(n-1)$ of these. We also find $n$ operators corresponding to the quasiprimary $\Oc= (TW) - {3\over 14}\p^2 W$ living on any of $n$ sheets, where $(TW)$ denotes the normal ordered product (e.g. \BouwknegtWG, p.33). All of the aforementioned are obviously quasiprimaries of $\C^n$, but none of them appears in the $\Phi_+\Phi_-$ OPE, on account of $\la W\ra_{\IC} = \la TW\ra_{\IC}=0$. Finally at level six, we see the ${n(n-1)\over 2}$ operators comprised of two $W$ currents that were analyzed in Section 4.3 and 4.4:
\eqn\chp{\Och_{W_{j_1}W_{j_2}} = W(z_{j_1})\otimes W(z_{j_2})}
In modes, \chp\ is just $W_{-3}$ acting on the vacuum of the $j_1$ and $j_2$ sheets. There are also other quasiprimaries at level six that do not contribute. 

This easily generalizes to the case of any spin-$s$ current living in an arbitrary higher spin algebra: we simply replace $\chi_3$ by the appropriate character.

\appendix{C}{R\'enyi entropy from $W$-conformal blocks?}
A natural starting point for the computation of twist field correlators in higher spin CFTs seems to be a conformal block decomposition under the full $W$ symmetry of the theory. This turns out to make things more complicated. 

The twist fields are $W$-primaries: they are annihilated by $W$-raising operators and are eigenfunctions of the $W$-zero modes,
\eqn\wcb{\eqalign{J^{(s)}_n|\Phi_{\pm}\ra&=0~, ~~ n>0\cr J^{(s)}_0|\Phi_{\pm}\ra &= Q_{\Phi_{\pm}}^{(s)}|\Phi_{\pm}\ra}}
However, one generically cannot, for instance, write a four-point function of $W$-primaries as a sum over products of three-point functions of primaries \BowcockWQ. 

Despite this, we are entitled to ask how the twist field correlators specifically behave in the semiclassical limit, $c\rar\infty$. When the higher spin algebra is $W_N$, for example, the universal dynamics of the current sector are controlled by sl(N,$\IR$) Toda theory, with flat space action
\eqn\wcc{S_{\rm Toda} = {1\over 8\pi}\int d^2z\left((\p\vec{\phi})^2+\mu\sum_{j=1}^{N-1}e^{b(e_j,\vec{\phi})}\right)}
where $\vec{\phi}$ is an $N-1$ vector, $e_j$ are simple roots of sl(N,$\IR$), and $(\cdot,\cdot)$ denotes an inner product on the weight space. The central charge is 
\eqn\wcd{c=N-1+N(N^2-1)(b+b^{-1})^2}
The semiclassical $b\rar 0$ limit of correlation functions is, as in Liouville theory, controlled by a saddle point approximation, so
\eqn\wcd{\la \Phi_+(0)\Phi_-(z)\Phi_+(1)\Phi_-(\infty)\ra_{\C^n/\IZ_n} \approx \exp\left[-c( f_{\rm Toda}+\overline{f}_{\rm Toda})\right]}
where the semiclassical block $f_{\rm Toda}$ is determined by a regularized on-shell Toda action evaluated on some solution of the Toda equation \FateevAB. However, unlike in Liouville theory, for generic external states the correlator does not obey an ordinary differential equation that could be used to fix $f_{\rm Toda}$. (This is related to our previous point about $W$ conformal blocks \FateevAB.) In the case that the external operators have some degeneracy, the correlator {\it does} obey a differential equation, but it is of higher than second order \refs{\FateevGS, \FateevAB}. 

It would be interesting to understand this work's conclusions from the behavior of $f_{Toda}$ for the specific case of external twist operators $\Phi_{\pm}$. A relevant fact seems to be that $\Phi_{\pm}$ have vanishing higher spin charge, $Q_{\Phi_{\pm}}^{(s>2)}=0$. To see this, we follow the arguments of \CalabreseEU. On the one hand, $\la J^{(s)}(z)\Phi_+(z_1)\Phi_-(z_2)\ra_{\IC} \propto \la J^{(s)}(z)\ra_{\IC} =0$, where the correlator on the left is evaluated in the $\C^n/\IZ_n$ theory, and we have used that $J^{(s)}$ is Virasoro primary. On the other hand, conformal invariance together with \wcb\ imply that $\la  J^{(s)}(z)\Phi_+(z_1)\Phi_-(z_2)\ra_{\IC}\propto Q_{\Phi_{\pm}}^{(s)}$ times a fixed function of coordinates. Thus, $Q_{\Phi_{\pm}}^{(s)}=0$. This is intimately related to the truncation of the bulk problem to the pure gravity sector.

\listrefs
\end